\documentclass[twoside,11pt]{article}

\usepackage[accepted]{melba}

\usepackage{amsmath,amsfonts}

\melbaheading{010}{https://www.melba-journal.org/article/27645}{2021}{1-23}{10/2020}{09/2021}{Shaw, Sudre, Ourselin, Cardoso and Pemberton}{Medical Imaging with Deep Learning (MIDL) 2020}{Marleen de Bruijne, Tal Arbel, Ismail Ben Ayed, Hervé Lombaert}

\ShortHeadings{A Decoupled Uncertainty Model for MRI Segmentation Quality Estimation}{Shaw, Sudre, Ourselin, Cardoso and Pemberton}
\firstpageno{1}

\title{A Decoupled Uncertainty Model \\ for MRI Segmentation Quality Estimation}

\author{\name Richard Shaw \email richard.shaw.17@ucl.ac.uk \\  
	\addr Dept. Medical Physics \& Biomedical Engineering, University College London, UK
	\AND
	\name Carole H. Sudre \email c.sudre@ucl.ac.uk
	\AND
	\name S\'{e}bastien Ourselin \email sebastien.ourselin@kcl.ac.uk
	\AND
	\name M. Jorge Cardoso \email m.jorge.cardoso@kcl.ac.uk \\
	\addr School of Biomedical Engineering \& Imaging Sciences, King’s College London, UK
	\AND
	\name Hugh G. Pemberton \email h.pemberton@ucl.ac.uk \\
	\addr Queen Square Institute of Neurology, University College London, UK
}

\begin{document}

\maketitle

\begin{abstract}
Quality control (QC) of MR images is essential to ensure that downstream analyses such as segmentation can be performed successfully. Currently, QC is predominantly performed visually and subjectively, at significant time and operator cost. We aim to automate the process using a probabilistic network that estimates segmentation uncertainty through a heteroscedastic noise model, providing a measure of task-specific quality. By augmenting training images with k-space artefacts, we propose a novel CNN architecture to decouple sources of uncertainty related to the task and different k-space artefacts in a self-supervised manner. This enables the prediction of separate uncertainties for different types of data degradation. While the uncertainty predictions reflect the presence and severity of artefacts, the network provides more robust and generalisable segmentation predictions given the quality of the data. We show that models trained with artefact augmentation provide informative measures of uncertainty on both simulated artefacts and problematic real-world images identified by human-raters, both qualitatively and quantitatively in the form of error bars on volume measurements. Relating artefact uncertainty to segmentation Dice scores, we observe that our uncertainty predictions provide a better estimate of MRI quality from the point of view of the task (gray matter segmentation) compared to commonly used metrics of quality including signal-to-noise ratio (SNR) and contrast-to-noise ratio (CNR), hence providing a real-time quality metric indicative of segmentation quality.
\end{abstract}


\section{Introduction}

Quality control (QC) in magnetic resonance imaging (MRI) is the process of establishing whether a scan or dataset meets a required set of standards. QC typically relates to the acceptable level of image quality required for a particular task, which may be affected by acquisition noise, resolution, and/or image artefacts induced for instance by blood flow, patient motion, field-of-view (FOV)/aliasing, bias field, and zipper or radio-frequency (RF) spikes. Indeed, in MRI, a large variety of potential artefacts need to be identified so that problematic images can either be excluded or accounted for in further image processing and analysis. To date, the gold standard for identification of these issues remains labour-intensive manual visual inspection of the data \citep{Graham2018}.

However, with the current trend towards acquiring and exploiting large imaging datasets, the time and resources required to perform visual QC have become prohibitive. Furthermore, 
visual QC is subject to inter and intra-rater variability due to differences in radiological training, rater competence, and sample appearance \citep{Sudre19}. Some artefacts, such as those caused by motion, can be difficult to detect with visual QC, as their identification requires careful examination of every slice in a volume. These challenges have led to increased interest in automated methods. In addition to the challenges inherent to visual QC, it is worth highlighting the task-dependent nature of a quality assessment: what is deemed acceptable quality for a radiological assessment may not be sufficient to provide reliable measurements for any automated analyses the image might undergo. Furthermore, manual QC is typically based on the perceptual ``visual quality’’ of the image rather than the acceptable level of quality required for a particular algorithmic task, such as segmentation. 

\begin{figure}[!tbp]
  \centering
  \begin{tabular}{c @{\qquad} c}
    \includegraphics[width=.3\linewidth]{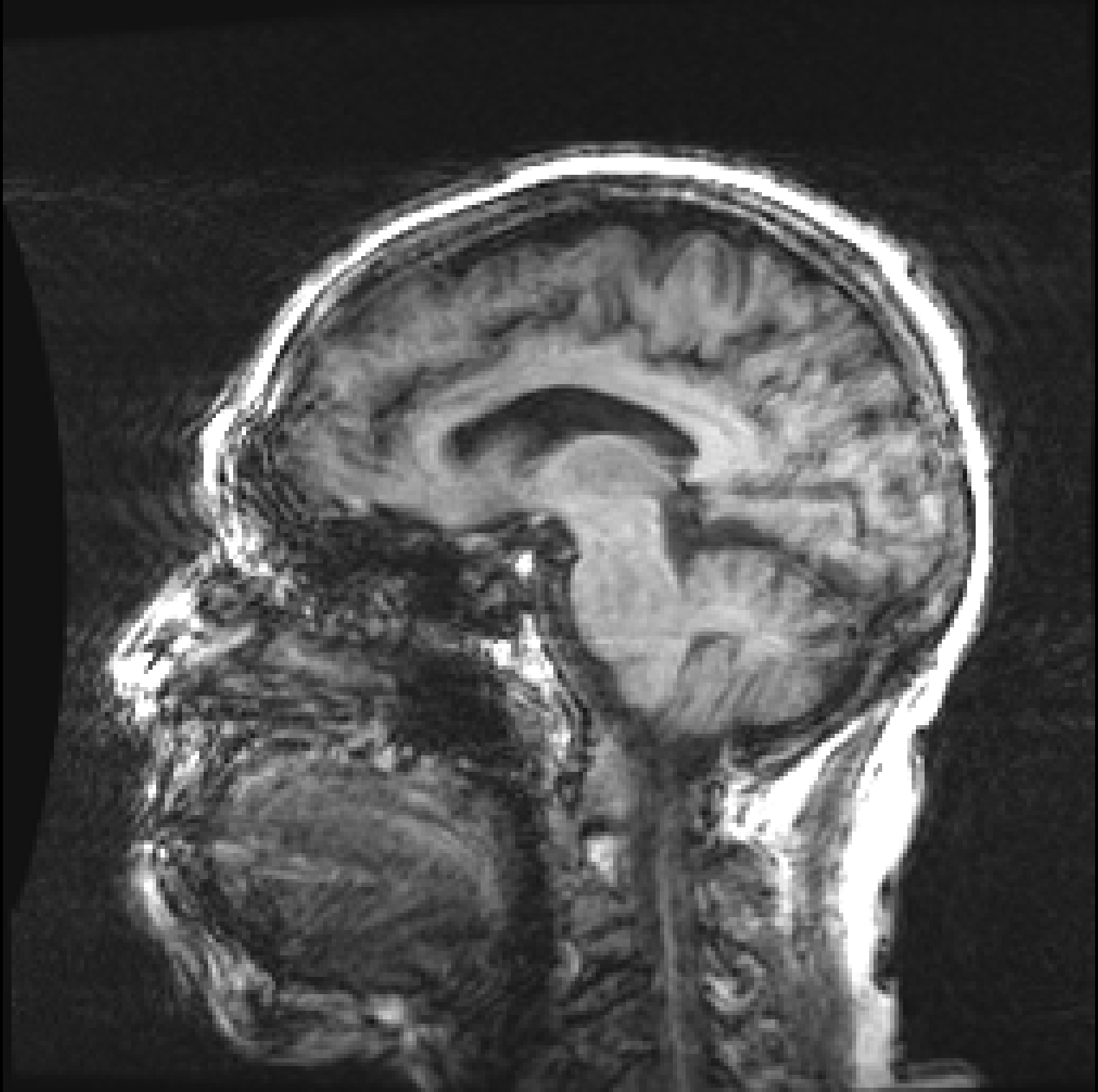} &
    \includegraphics[width=.3\linewidth]{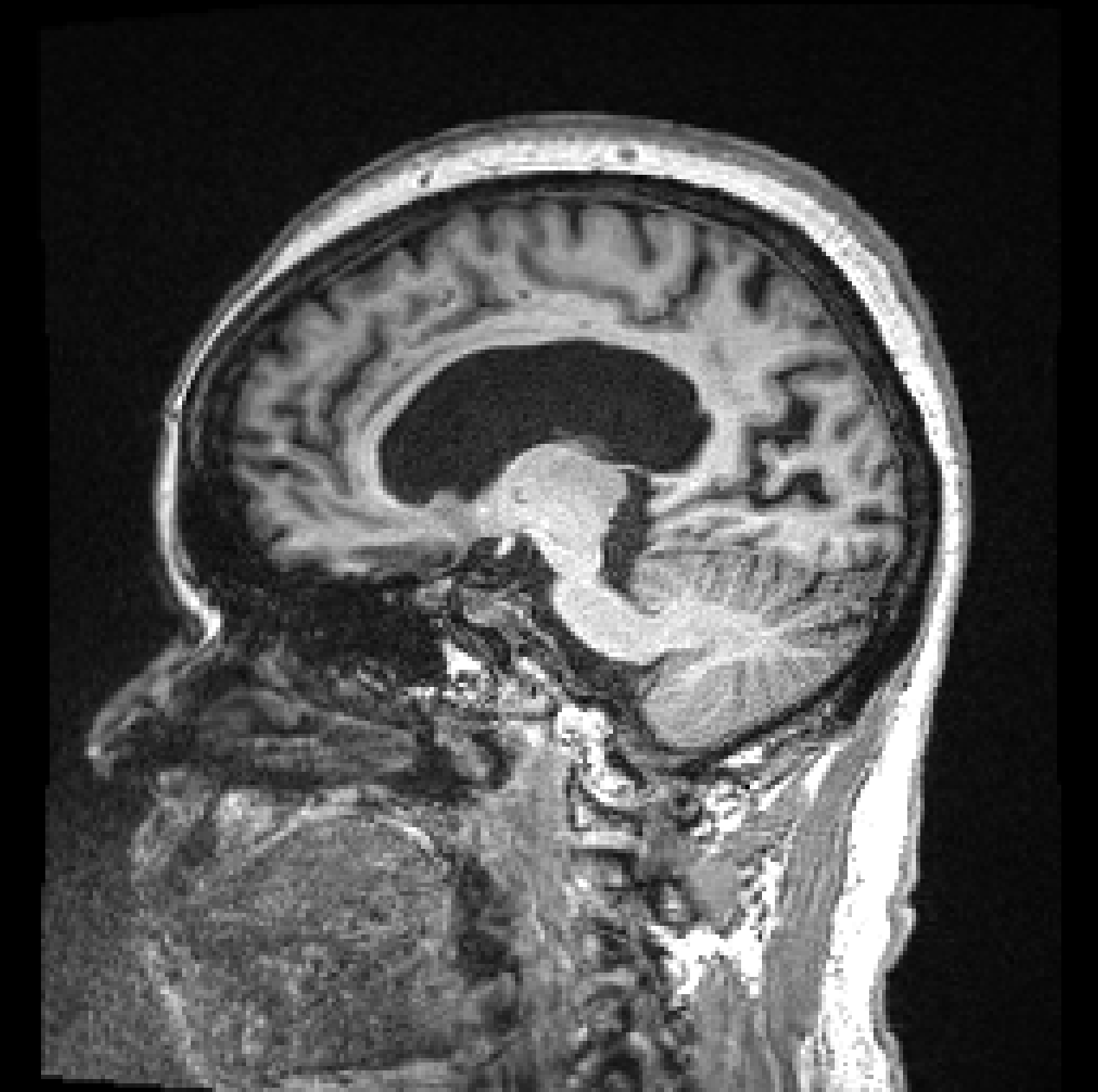} \\
    \small (a) & \small (b)
  \end{tabular}
  \caption{Two scans failing visual QC due to motion artefacts. a) The artefact located in the brain and affecting downstream gray matter segmentation task. b) Artefact localised at the bottom of the image, leaving the brain unaffected. From an algorithmic point of view, this image could have been used for a segmentation task, highlighting the distinction between visual and algorithmic QC.}
  \label{fig:qc}
\end{figure}

 Figure \ref{fig:qc}, presents two T1 images from the ADNI (Alzheimer's Disease Neuroimaging Initiative) database, illustrating this difference. Both images, graded by trained analysts, have been deemed ``unusable" for containing artefacts due to motion (the criteria used for the QC is described in detail on the ADNI website\footnote[1]{\url{http://adni.loni.usc.edu/methods/mri-tool/mri-analysis/}}). However, since in the right image the artefacts do not intersect the brain, the task of segmenting gray matter would not be affected and the scan should have been kept for this analysis. Thus, there is a distinction between perceptual quality 
 and the quality required for algorithmic processing. Furthermore, by training with 
 augmentation, convolutional neural networks (CNNs) make the internal representation of the data more robust to noise, so from an algorithmic perspective, some image degradation may be tolerated before observing a 
 drop in performance. These observations allow us to define MRI quality 
 as the model's ability to perform the task, rather than a subjective visual grading of quality. 
 
 Therefore, in this work, we propose to estimate uncertainty using a Bayesian Deep Learning framework and show that the resulting uncertainty predictions can be used as a measure of quality for the task of gray matter segmentation. Furthermore, to isolate the impact of artefacts on the segmentation, we present a novel decoupled uncertainty model, where 
 decoupled uncertainty predictions reflect 
 the presence of different artefacts, assuming that the artefact processes are conditionally independent. For example, patient movement and RF-spikes are clearly independent effects. Furthermore, 
 in terms of the acquisition physics, the point spread function (PSF) is dependent on image resolution and window size, while noise is a product of the receiver coils, meaning that blurriness due to low resolution is independent of the coil noise and sensitivity. 
 
 The ability to 
 identify sources of uncertainty can have a direct impact on the management of both clinical and research logistics. For instance, if the observed uncertainty is associated with acquisition artefacts (noise, for example) inspection of the scanner by an engineer may be required. In contrast, if the uncertainty stems from subject motion, 
 recall of the subject may be the appropriate path of action. In addition, in the context of population studies, obtaining the uncertainty associated with the desired measurement allows for appropriate statistical treatment of the samples while limiting the number of exclusions for quality reasons. Lastly, real-time indication of levels of uncertainty concerning a downstream task would enable radiographers to best manage session time and ensure repeat scans are taken as needed.
 
Given the challenge of decoupling the effect of different artefacts, we must appreciate that the image degradation possibilities in MRI are wide-ranging and artefact appearance is varied. Artefacts may be image-wide or localised within a volume, making their detection difficult. Classification of artefacts and attributing labels to artefact sub-types is also problematic. For instance, at a voxel-level, patient motion can appear similar to blurring or Gibbs ringing. Furthermore, motion artefacts are often categorised by QC-raters into ``blurring,'' ``ringing'' and ``ghosting,'' but it can be difficult to distinguish between them. Artefact classification is not consistent across raters and is protocol dependent; for example, artefacts may be ignored if they lie outside of the skull. These QC inconsistencies impede the use of real-world QC labels for training/validating of models. Additionally, multiple artefact sub-types may be present in a scan, further complicating the decoupling task. However, 
modelling the noise due to artefacts through an uncertainty model allows us to handle this naturally. Our proposed probabilistic network, discussed in section 3, learns the best way it can (from the point of view of minimising the loss) of decoupling uncertainty from the data in a self-supervised fashion, without the need for explicit artefact labels.
 \\
 
 \noindent \textbf{Contributions} The main contributions of this work are three-fold:
\begin{enumerate}
    \itemsep0em 
    \item A general method of estimating MRI segmentation quality in a self-supervised manner.
    \item A novel cascading student-teacher CNN architecture and probabilistic loss function to decouple sources of uncertainty related to the task and different image artefacts.
    \item Validation of uncertainty predictions as a measure of segmentation quality on problematic images identified by expert raters.
\end{enumerate}

 \noindent \textbf{Journal Extensions} This manuscript builds upon our work presented at the MIDL 2020 conference \citep{Shaw2020AHU}. In this paper we make the following extensions:
\begin{enumerate}
    \itemsep0em 
    \item Better motivation of the problem is described; specifically, the issue of ``visual'' QC vs ``algorithmic'' QC with regards to segmentation quality. We also discuss the advantages of being able to decouple predictive uncertainty due to artefacts and why, for our model, specific sources of artefacts may be considered independent.
    
    \item A more extensive literature review to reflect the growing research into automated QC using Deep Learning and uncertainty-based methods. Since our original paper was published, a number of papers have been released tackling automatic QC and, in a similar approach to our work, using uncertainty for this purpose. Due to the paper's extended length, the related works section is now divided into sub-sections for clarity.
    
    \item Our model has been extended to include patient movement artefact augmentation. Prior work did not include motion artefacts due to time limitations. This was a significant barrier to validating on real-world data since motion artefacts are one of the most common types of MRI artefacts in the datasets we have available.
    
    \item Extended quantitative validation experiments have been carried out to investigate the use of decoupled artefact uncertainty. In these additional experiments,  we show that by relating the decoupled artefact uncertainty from our model to segmentation  Dice scores, the uncertainty predictions can be used as a metric of segmentation quality. We compare this metric to signal-to-noise ratio (SNR) and contrast-to-noise ratio (CNR) and show that on both simulated data and real-world data, our metric of quality is better correlated to the segmentation quality as measured by Dice score.
    
    \item The discussion and limitations sections have been extended to address comments from reviewers. In short, the model has shown promising results, but we want to highlight a number of current limitations with the model that readers should be made aware of if the work is to be taken forwards for future research.
\end{enumerate}

\section{Related Works}

\noindent \textbf{Uncertainty in Deep Neural Networks}
The basis of this work uses Bayesian 
CNNs to predict the uncertainty of segmentation predictions. In recent years, estimating uncertainty in the data/model has become increasingly recognised as an important step to enable the safe transition of automated methods into the clinical environment, for instance in \citep{Wang2018}, \citep{Kohl2018APU}, \citep{Tanno2019UncertaintyQI} and \citep{Hu2019Supervised}. In Bayesian Deep Learning, two main types of uncertainty are commonly distinguished:  \textit{epistemic} uncertainty which is uncertainty in the model, and \textit{aleatoric} uncertainty which depends on noise or randomness in the data \citep{Kendall2}. 

Epistemic uncertainty is modelled by placing a prior distribution over the model weights and capturing how the weights vary given the data. Often, Bayesian inference is computationally intractable and is approximated using Laplace approximation, Markov chain Monte Carlo methods or variational methods \citep{Combalia2020Uncertainty}. Monte Carlo dropout is used to create an ensemble of predictions, and the uncertainty is estimated as the variance over predictions. It consists of training a model with dropout at each layer of the network and performing many forward passes of the dropout network to sample from the approximate posterior \citep{Gal2016Dropout}. Alternatively, predictions can be generated using an ensemble of differently trained neural networks \citep{Lak2017}.

Aleatoric uncertainty, on the other hand, captures noise inherent in the observations. This could be a result of measurement noise, resulting in uncertainty which cannot be reduced even with the collection of more data. To capture the aleatoric uncertainty, we learn the observation noise parameter $\sigma$, 
by minimising a log-likelihood loss function assuming that the noise can be modelled with some distribution - usually, a Normal distribution is assumed, as we do in this work. In a similar approach to the one presented in this work, \cite{Prado2019DualNN} have used a dual network to learn both epistemic and aleatoric uncertainty, assuming their independence. However, since the focus of this work is on the assessment of image quality, only the aleatoric uncertainty is considered here. \newline

\noindent \textbf{Deep Learning-based QC}
Estimating MRI quality for automated QC remains an open problem, but a number of learning-based approaches have been proposed. \cite{Muelly} 
use CNNs for three QC tasks: 1) classifying MRI diagnostic quality, 2) determining if artefacts are present, and 3) classifying scan sequence type. They use 
transfer learning on a dataset of images categorised by an experienced radiologist. In \cite{Graham2018}, the authors use CNNs to detect motion artefacts in diffusion-weighted MRI. Addressing the drawback of supervised approaches, which require time-consuming and subjective manually labelled data for training, simulated data is used in addition to some manually labelled real-world data for a final calibration step. 

Using a GAN, \cite{Brusini2020ADL} estimate the quality of brain segmentations by learning a mapping from the segmentation to the MR image with the pix2pix framework \citep{Isola2017ImagetoImageTW}. Comparing the generated MRI from a potentially bad segmentation to the original MRI results in an error map, highlighting regions where they do not match, i.e. where segmentation errors may be present. Also using a generative model, \cite{Wang2020DeepGM} learn a manifold of ``good-quality'' image segmentation pairs of Cardiac MRI scans. The quality of a given test segmentation is assessed by evaluating the difference from its projection onto the good-quality manifold. 

With a similar approach to ours, estimating image quality from 
segmentation uncertainty, \cite{Devries2018LeveragingUE}  propose a two-stage architecture. 
First, a CNN produces a segmentation prediction and uncertainty map, and second, a segmentation quality CNN produces a quality estimate. The uncertainty map can be used to interpret the segmentation network’s output, while the quality estimate can automatically reject, or alert the user to poor segmentations. 
However, our work remains distinct through the use of novel k-space augmentations to isolate uncertainty due to image artefacts.
Also using segmentation uncertainty of T1 images for QC purposes, \cite{Roy2019BayesianQM} introduce Bayesian QuickNAT. They predict epistemic, or model, uncertainty utilising test-time dropout sampling to generate multiple segmentation realisations, and introduce four metrics to measure the structure-wise uncertainty. They show that by increasing input image noise the uncertainty is correlated with segmentation Dice score. However, our work differs from this by modelling the aleatoric uncertainty with a heteroscedastic model allowing us to learn the data-dependent uncertainty as a function of the input data.
\newline

\noindent \textbf{Classical QC Methods}
Non-Deep Learning approaches to MRI QC include \cite{Sun}, an open-source tool performing a number of QC tests including: geometric distortion, slice thickness/position accuracy, percentage signal ghosting, high contrast spatial resolution and low contrast object detectability. \cite{Elias} present an automated method for assessing MRI quality acquired in a clinical trials environment, with the following QC procedures: 1) patient identity verification, 2) alphanumeric parameter matching, 3) SNR estimation, 4) gadolinium-enhancement verification, and 5) detection of head movement ghosting. Each procedure provides an automated quantitative measurement which is compared against an acceptance threshold based on receiver operating characteristics of traditional manual and visual QC performed by trained experts. \cite{Sadri2020MRQyAO} present an open-source QC tool MRQy with two main objectives: 1) to interrogate MRI cohorts for site- or equipment-based differences, and 2) quantify the impact of MRI artefacts on relative image quality. MRQy extracts several quality measures: noise ratios, variation metrics, entropy and energy criteria, and MR image metadata.
\cite{Esteban2017MRIQCAT} show that task-dependent measures such as entropy-focus criterion (EFC) and INU (intensity non-uniformity) amongst others, can be used for QC, targeting specific artefact sub-types.
\newline

\noindent \textbf{Data Augmentation}
In Deep Learning, data augmentation is important to making networks more generalisable and robust to image corruption/unseen data. The amount and type of augmentation to apply during training is an important aspect of this work and learning the best polices of data augmentation remains an active area of research \citep{RandAugment}. Image augmentations often include geometric transforms such as scaling, translation, rotation, flipping, and elastic deformations, or changes to image intensity, gamma, saturation, and random noise. More recently, augmentation methods blending multiple images/augmentations together such as AugMix \citep{Hendrycks2020AugMixAS}, CutMix \citep{Yun2019CutMixRS}, and MixUp \citep{Zhang2018mixupBE}, have demonstrated improved model performance and robustness. Our augmentation pipeline is similar, except augmentations are combined in the k-space to better reflect the MRI acquisition process. 
Ideally, any data augmentation should be physically plausible for the particular task. As such, we use a k-space augmentation model to generate physically realistic artefacts for MR images, detailed in section 5. 
\newline

\noindent \textbf{Student-Teacher Networks}
Our decoupled uncertainty model (section 3) is based on 
knowledge distillation and student-teacher networks \citep{Hinton2015}. Distillation, usually implemented for 
model compression, uses a large network to teach a smaller one with fewer parameters. The large network learns ``soft'' labels 
and consistency losses 
force the smaller one to replicate it's behaviour. 
Here, the aim is not compression, but output decoupling: a student network learns the outputs of separate teacher networks trained with different augmentations. Each teacher learns the uncertainty due to a single artefact sub-type, and a student learns all uncertainties by attempting to replicate their respective outputs. This side-steps the need for labels of uncertainty, since the uncertainty is learned in an unsupervised way. 
Our work resemblances ``Noisy Student'' training \citep{Xie2020SelfTrainingWN}, which extends distillation with equal-or-larger student models. A teacher generates ``pseudo labels'' for unlabelled images and a student is trained on both labelled and pseudo labelled images. The process is iterated replacing the student with the teacher, and injected noise (dropout/augmentation) enables the student to generalise better than the teacher. 

\section{A Heteroscedastic Aleatoric Uncertainty Model}
In this section we present an uncertainty model to predict aleatoric uncertainty. The aleatoric uncertainty is classically divided into two categories; the \textit{homoscedastic} component is the task-dependent uncertainty, while the \textit{heteroscedastic} component depends on the input data, reflecting for instance its quality, and can be predicted as a model output.
Following this classification, task-specific image quality is modelled according to a heteroscedastic noise model. Heteroscedastic models assume that observation noise $\sigma^2$ can vary with the input $\mathbf{x}$, allowing for regions of the observation space to have higher noise levels than others \citep{Kendall2}.

In this work, a single task, grey matter segmentation is considered; thus task uncertainty should be similar across experiments.  The total predicted uncertainty is further assumed to be the sum of the task uncertainty (uncertainty given clean data) and the heteroscedastic uncertainties introduced as a function of image corruption.

For the segmentation task,  the problem is presented as a voxel-wise classification.
The likelihood to maximise is defined as the softmax function of the scaled output logits, i.e. $p(\mathbf{y} | \mathbf{f^W}(\mathbf{x}), \sigma) = \mathrm{Softmax} \left( \mathbf{f^W}(\mathbf{x}) / \sigma^2 \right)$ where $\mathbf{f^W}(\mathbf{x})$ is the output of a neural network with weights $\mathbf{W}$ and input $\mathbf{x}$ \citep{Bragman2018}.
The negative log likelihood is therefore:

\begin{align}
    -\log p(\mathbf{y} = c | \mathbf{f^W}(\mathbf{x}), \sigma) 
    &=
    -\log \mathrm{Softmax} \left( \frac{1}{\sigma^2} \mathbf{f}_c^{\mathbf{W}}(\mathbf{x}) \right) \\
    &=
    -\frac{1}{\sigma^2} \mathbf{f}_c^{\mathbf{W}}(\mathbf{x}) 
    +
    \log \sum_{c'} \exp
    \left(
    \frac{1}{\sigma^2} \mathbf{f}_{c'}^{\mathbf{W}}(\mathbf{x})
    \right)
\end{align}

\noindent where $\mathbf{f}_c^{\mathbf{W}}(\mathbf{x})$ is the $c$\textsuperscript{th} element of the output vector $\mathbf{f^W}(\mathbf{x})$.
Note, in practice for segmentation we compute the unscaled cross entropy loss of $\mathbf{y}$, given by $\mathrm{CE}\left(
    \mathbf{y} = c, \mathbf{f^W}(\mathbf{x})
    \right) 
    =
    -\log \mathrm{Softmax} \left(
    \mathbf{f}_c^{\mathbf{W}}(\mathbf{x}) \right)
    =
    -\mathbf{f}_c^{\mathbf{W}}(\mathbf{x}) 
    +
    \log \sum_{c'} \exp
    \left(
    \mathbf{f}_{c'}^{\mathbf{W}}(\mathbf{x})
    \right)$. Substituting this into Eq. 2:

\begin{equation}
  \begin{aligned}
  -\log p(\mathbf{y} = c | \mathbf{f^W}(\mathbf{x}), \sigma) 
    &=
    \frac{1}{\sigma^2} \mathrm{CE} 
    \left(
    \mathbf{y} = c,\mathbf{f}^{\mathbf{W}}(\mathbf{x})
    \right) 
    +
    \log 
    \frac{\sum_{c'} \exp
    \left(
    \frac{1}{\sigma^2} \mathbf{f}_{c'}^{\mathbf{W}}(\mathbf{x})
    \right)}{\left(
    \sum_{c'} \exp
    \left(
    \mathbf{f}_{c'}^{\mathbf{W}}(\mathbf{x})
    \right)
    \right)^{\frac{1}{\sigma^2}}}
 \end{aligned}
\end{equation}

Following \cite{Kendall2017}  the likelihood is approximated: 
$\left(
    \sum_{c'} \exp
    \left(
    \mathbf{f}_{c'}^{\mathbf{W}}(\mathbf{x})
    \right)
    \right)^{\frac{1}{\sigma^2}} \approx 
    \frac{1}{\sigma}
    \sum_{c'} \exp 
    \left(
    \frac{1}{\sigma^2}
    \mathbf{f}_{c'}^{\mathbf{W}}(\mathbf{x})
    \right)$. Substituting into Eq. 3 results in the weighted cross entropy loss that is used as the base loss function for all segmentation networks, as given by Eq. \ref{eq:loss}.

\begin{equation}
    \mathcal{L}_{NN} =
    \frac{1}{\sigma^2 }\mathrm{CE}\left(
    \mathbf{y},\mathbf{f}^{\mathbf{W}}(\mathbf{x})
    \right)
    + \frac{1}{2}
    \log \sigma^2
    \label{eq:loss}
\end{equation}

\section{Decoupling Multiple Uncertainties} 
\begin{figure}[t!]
    \centering
    \includegraphics[width=1.0\textwidth]{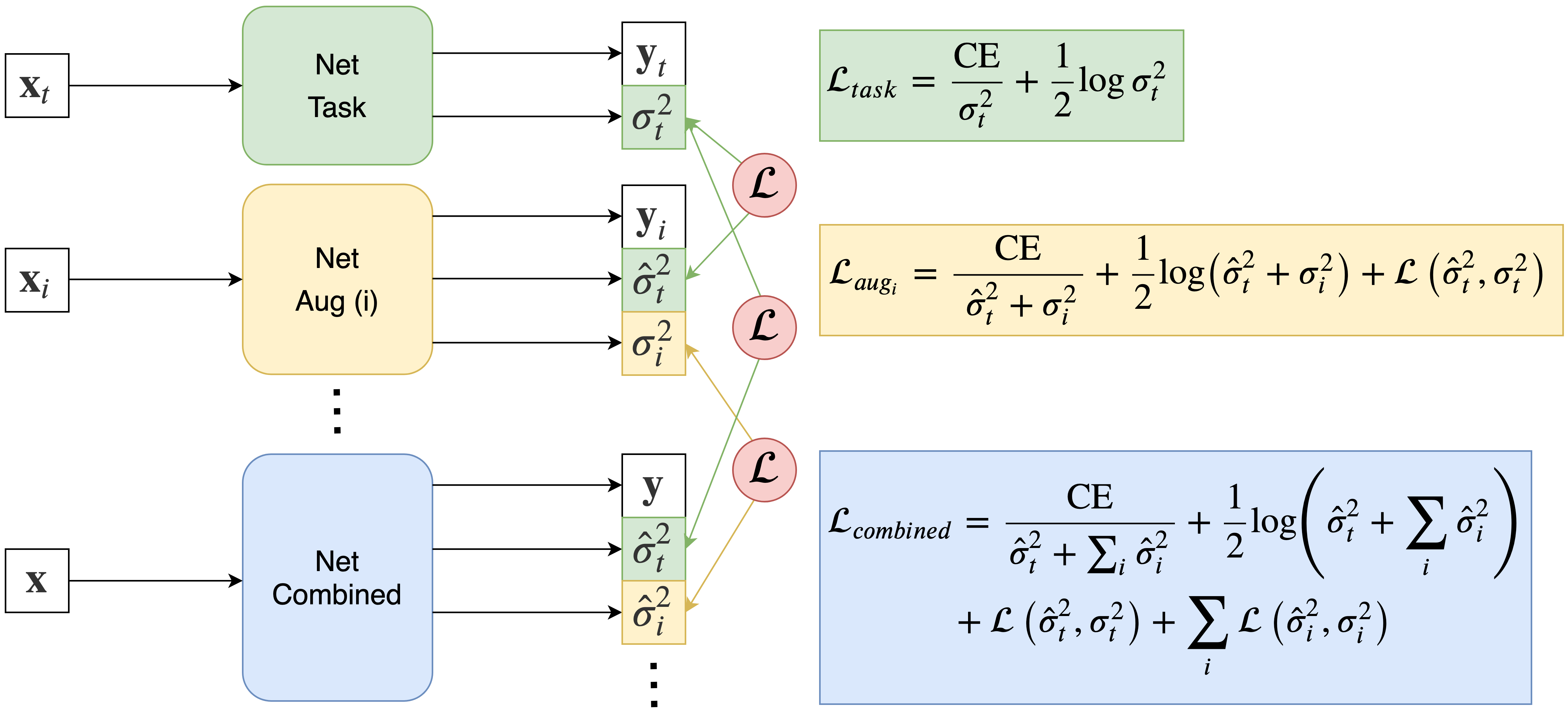}
     \caption{Proposed network architecture and training strategy: First, the task network is trained on clean images $\mathbf{x}_t$ to predict segmentation $\mathbf{y}_t$ and task uncertainty $\sigma_t^2$. Then, for each augmentation $i$, a teacher network is trained to predict both the task uncertainty $\hat{\sigma}_t^2$ and augmentation uncertainty $\sigma_i^2$, where the output from the task network supervises with consistency loss $\mathcal{L}(\hat{\sigma}_t^2, \sigma_t^2)$. Lastly, a combined student network is trained, where the uncertainty outputs from all previous teacher networks supervise its uncertainty predictions in a similar fashion. The loss functions for each CNN at each training stage are shown in corresponding colours.}
    \label{fig:network1}
\end{figure}

We use the weighted cross entropy loss in Eq. \ref{eq:loss} to learn voxel-wise uncertainty, i.e. the network has two outputs: the segmentation $\mathbf{y}$ and the variance $\sigma^2$, as shown by the task network in Figure \ref{fig:network1}. We adapt this loss function to predict multiple uncertainty quantities related to different aspects of image quality. The aim is to decompose the total predicted uncertainty $\sigma^2$ into multiple uncertainty quantities related to the inherent difficulty of the task and to different types of image degradation or augmentation that may affect image quality. Our model assumes the variance sum law for independent events, such that the total predicted variance is the sum of the individual variances associated with each mode of augmentation, i.e. $\sigma^2 = \sigma_t^2 + \sigma_1^2 + \sigma_2^2 + ... + \sigma_N^2 = \sigma_t^2 + \sum_{i=1}^{N} \sigma_i^2$ for $N$ possible augmentations, where $\sigma_t^2$ is the task uncertainty and $\sigma_i^2$ is the uncertainty due to the $i$\textsuperscript{th} augmentation. This assumption of independence has the merit to simplify the model and ensure training tractability. While interactions with task uncertainty (task harder to learn with noisier data) or between degradation types (blurring and noise for instance) exist, their modelling would require the learning of new covariance terms and would greatly complexify both model and training procedure.

Substituting for $\sigma^2$ in Eq. \ref{eq:loss} results in the combined loss function in Eq. \ref{eq:combinedloss}.

\begin{equation}
    \mathcal{L}_{combined} =
    \frac{\mathrm{CE} \left(
    \mathbf{y}, \mathbf{f^W}(\mathbf{x})
    \right)}{\sigma_t^2 + \sum_{i=1}^{N} \sigma_i^2}
    + \frac{1}{2}
    \log \left(\sigma_t^2 + \sum_{i=1}^{N} \sigma_i^2 \right)
    \label{eq:combinedloss}
\end{equation}

The purpose of $\mathcal{L}_{combined}$ is to enable the network to predict task uncertainty $\sigma_t^2$ and augmentation uncertainty $\sigma_i^2$ for each mode of augmentation. Since networks trained with this probabilistic loss function learn uncertainty in an unsupervised way, we do not have explicit labels for uncertainty. Therefore the network cannot determine how to decompose the total variance into separate quantities $\sigma_t^2$ and $\sigma_i^2$ by itself, i.e. without supervision. To do this, inspired by student-teacher networks \citep{Tarvainen2017} and knowledge distillation \citep{Hinton2015}, \citep{NoisyStudent}, we use a series of intermediate teacher networks where each network predicts the uncertainty due to a single augmentation, creating ``pseudo labels'' of uncertainty maps. By training these teacher networks sequentially, the output uncertainties from each intermediate network are used as self-supervising labels for the uncertainties predicted by a final combined student network. This is shown schematically in Figure \ref{fig:network1}.

The training procedure can be summarised in the following three steps: 1) Train a teacher network on clean data only to predict the segmentation $\mathbf{y}_t$ and task uncertainty $\sigma_t^2$. 2) Freeze the task network and train a new network $\mathcal{N}_i$ for each mode of augmentation we wish to decouple, where each augmented network predicts the segmentation $\mathbf{y}_i$, the task uncertainty and the noise uncertainty $\sigma_i^2$ for augmentation $i$. The output uncertainty from the first network acts as a “pseudo label” for the task uncertainty. 3) Freeze all previous networks and train a final student network with all modes of data augmentation to predict the task uncertainty and all possible augmentation uncertainties, where each uncertainty is supervised by the pseudo uncertainty labels from their respective teacher networks. 

For each network to learn the task uncertainty, an additional consistency loss term $\mathcal{L}(\hat{\sigma_t}^2, \sigma_t^2)$ is added to the weighted cross entropy loss to minimise the difference between the uncertainty outputs from two networks. Therefore, each augmentation network $\mathcal{N}_i$ minimises a loss function $\mathcal{L}_{aug_i}$ given by Eq. \ref{eq:augloss}, 

\begin{equation}
    \mathcal{L}_{aug_i} =
    \frac{\mathrm{CE}
    \left(
    \mathbf{y}_i, \mathbf{f^W}(\mathbf{x})
    \right)}{\sigma_t^2 + \sigma_i^2}
    + \frac{1}{2}
    \log \left(\sigma_t^2 + \sigma_i^2 \right)
    + \mathcal{L}(\hat{\sigma_t}^2, \sigma_t^2)
    \label{eq:augloss}
\end{equation}

\noindent where, 
\begin{equation}
\mathcal{L}(\hat{\sigma}^2, \sigma^2) =
 \mathcal{L}_1(\hat{\sigma}^2, \sigma^2)
+ \mathcal{L}_{grad}(\hat{\sigma}^2, \sigma^2)
+ \lambda \mathcal{L}_{SSIM}(\hat{\sigma}^2, \sigma^2).
\end{equation}

$\mathcal{L}_1$ is the L1 loss of the uncertainty and $\mathcal{L}_{grad}$ is the L1 loss of gradient differences of the uncertainty maps in all three axes. The term $\mathcal{L}_{SSIM}$ computes the 3D structural similarity (SSIM). The gradient and structural similarity losses help preserve the structure of the predicted uncertainty maps as the level of degradation increases. However, in the presence of severe image artefacts, the position, shape, appearance/visibility of the segmentation boundary can change causing SSIM to breakdown. Therefore $\mathcal{L}_{SSIM}$ is down-weighted by $\lambda = 0.1$. A simplified SSIM with a $3 \times 3 \times 3$ average filter is used in our implementation.

This choice of consistency loss function was empirically driven. Using an L2 loss produced over-smoothed uncertainty estimates, while using L1 terms and structural similarity produced sharper uncertainty predictions. This particular form of loss terms is a common choice for many pixel-wise regressions problems, for example in \cite{Zhao2017LossFF}.

\section{k-Space Augmentation}

\begin{figure*}[!t]
  \centering
  \begin{tabular}{c @{\hspace{0.25em}} c @{\hspace{0.25em}} c @{\hspace{0.25em}} c @{\hspace{0.25em}} c}
    \includegraphics[width=.19\linewidth]{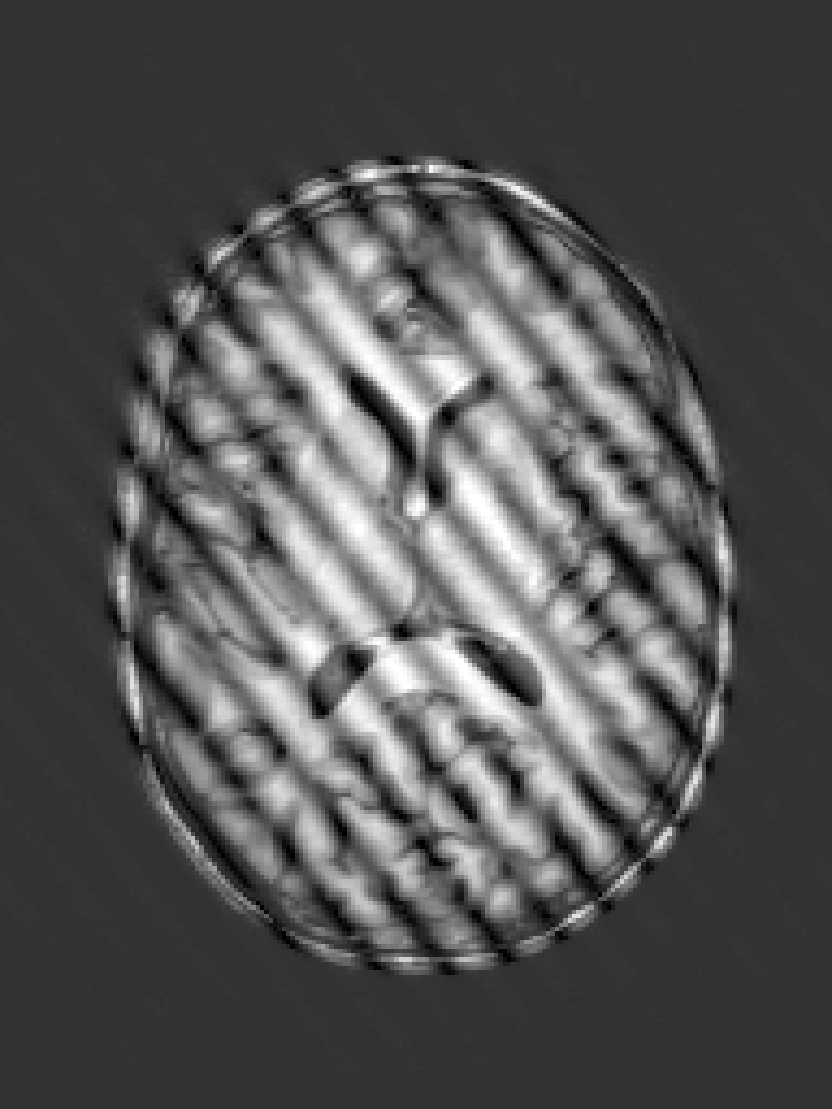} &
    \includegraphics[width=.19\linewidth]{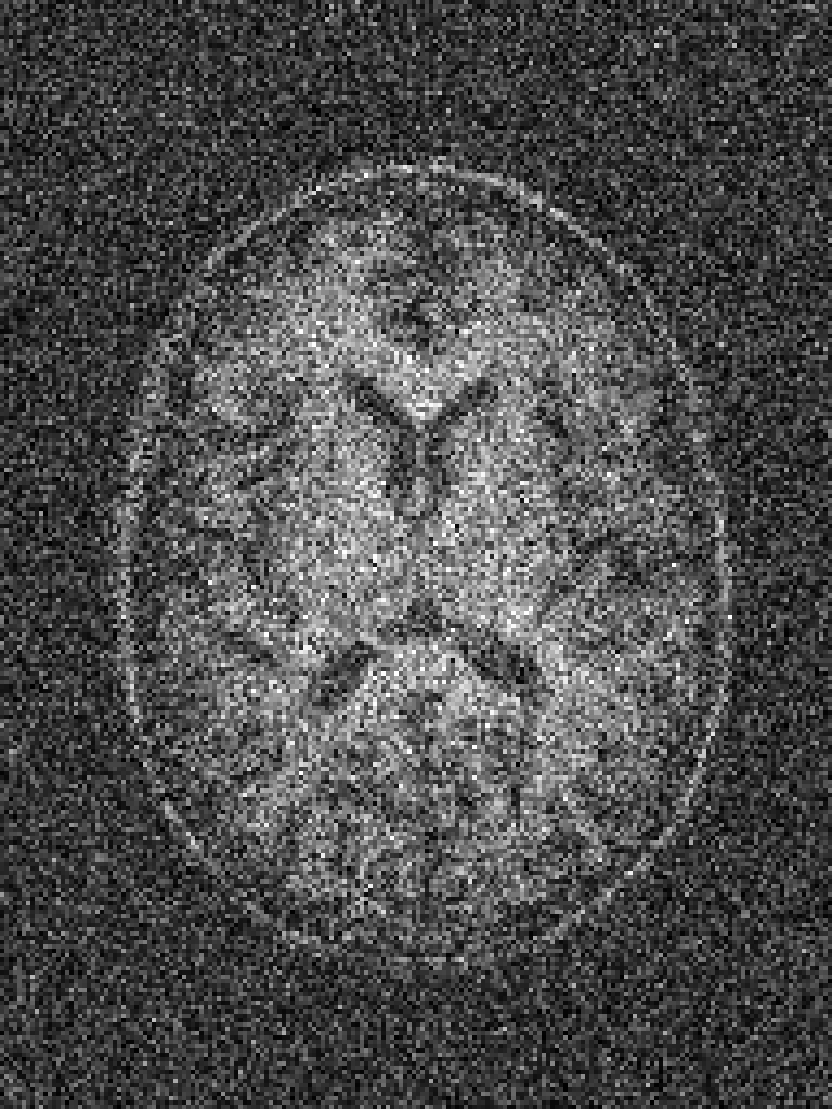} &
    \includegraphics[width=.19\linewidth]{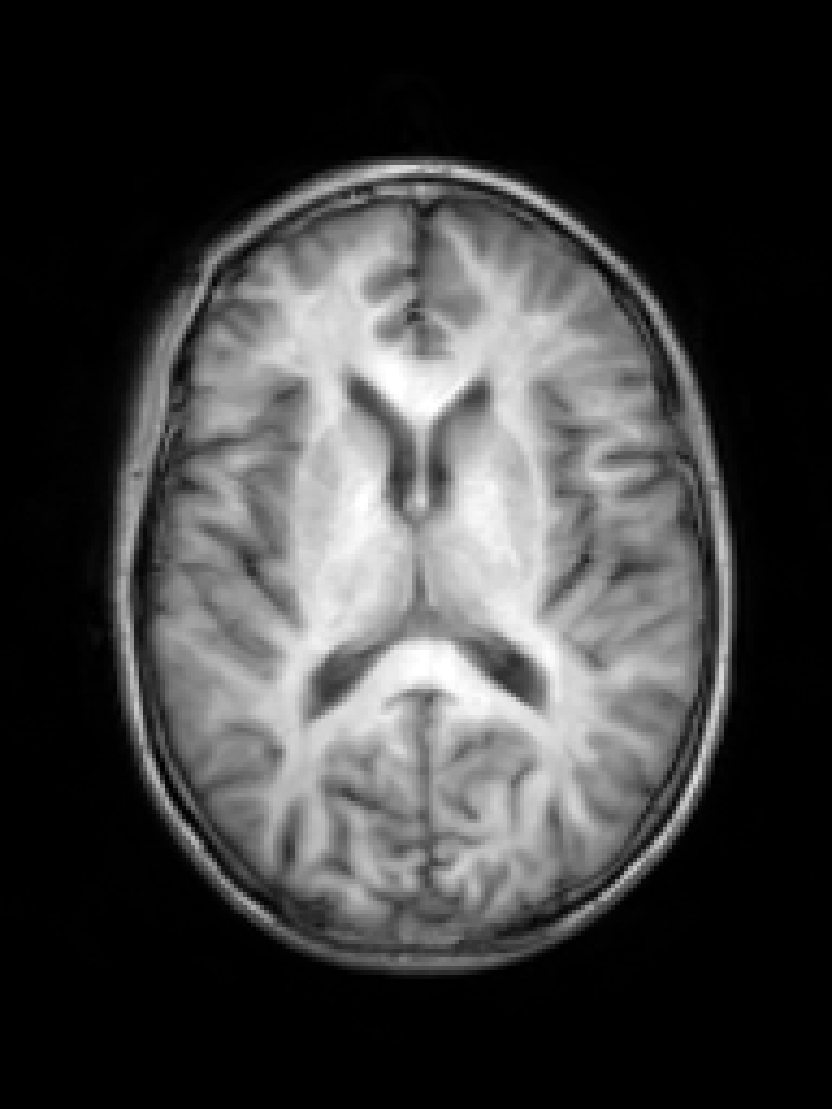} &
    \includegraphics[width=.19\linewidth]{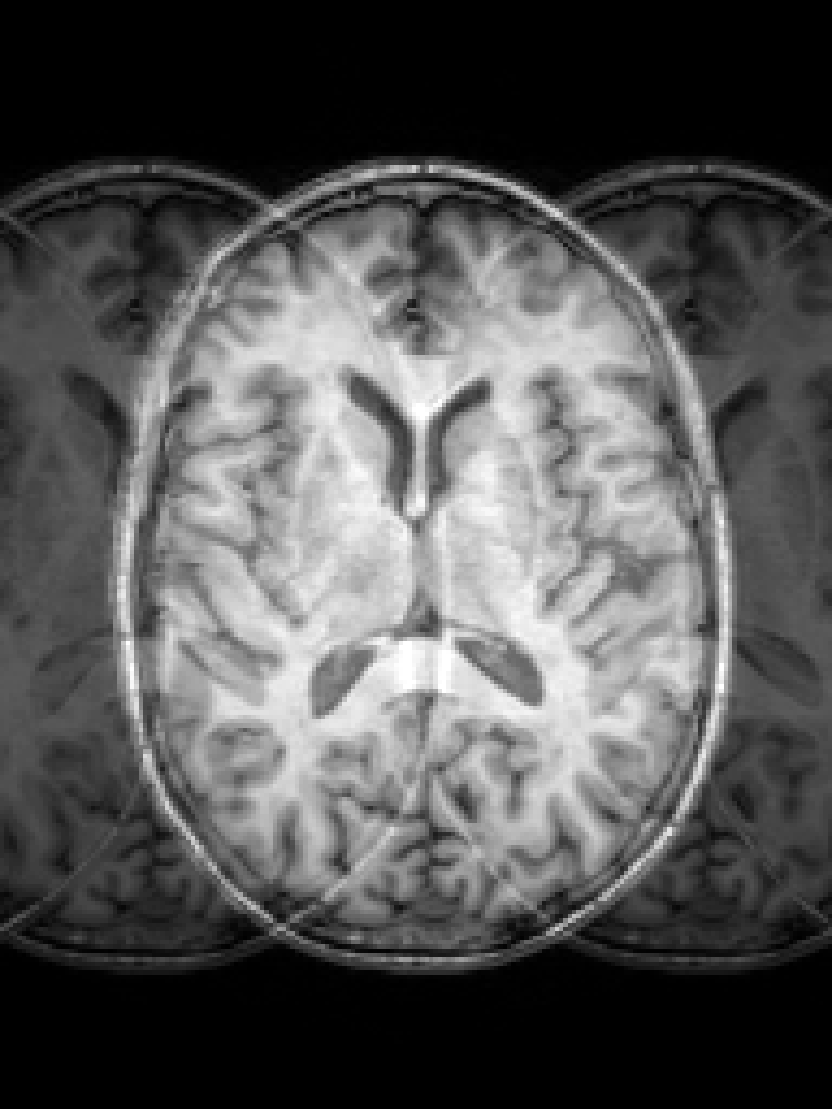} &
    \includegraphics[width=.19\linewidth]{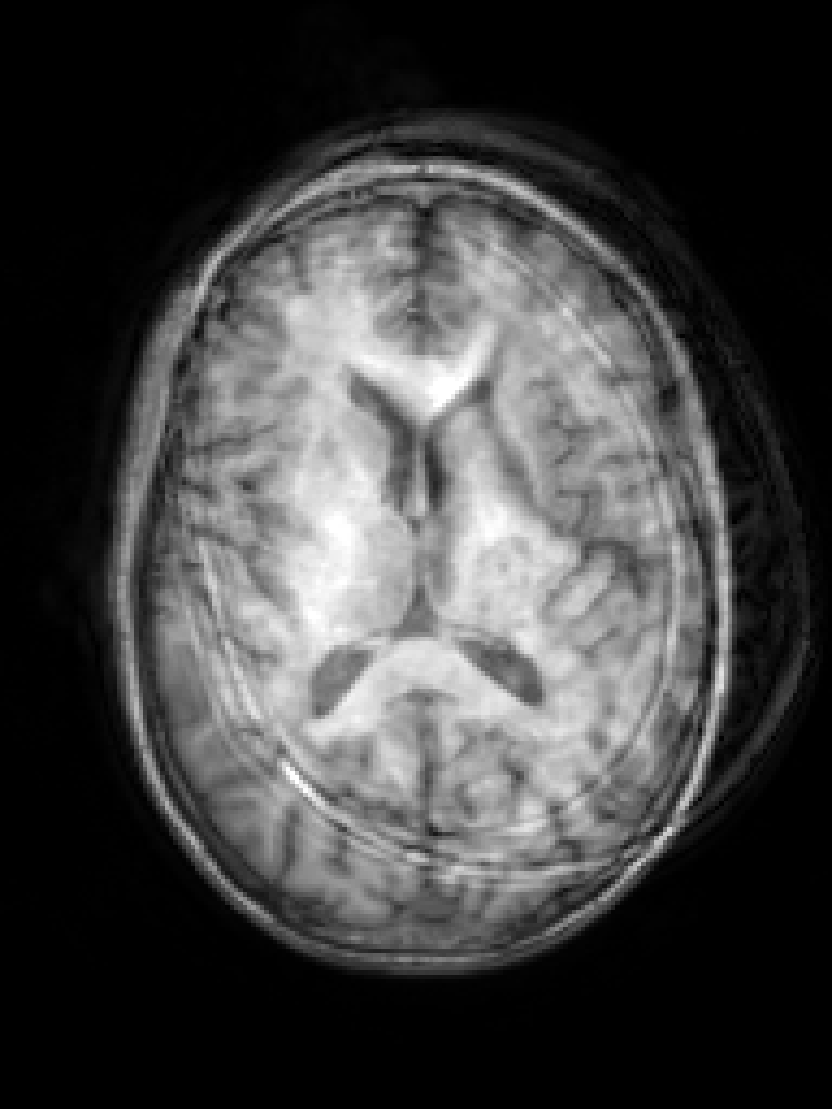} \\
    \small (a) & \small (b) & \small (c) & \small (d)  & \small (e)
  \end{tabular}
  \caption{Examples of simulated k-space augmentations applied during the training of each CNN, a) RF spike artefact b) Gaussian noise in the k-space c) low-pass filter along one or more axes d) aliasing/wraparound artefact e) patient movement artefact.}
  \label{fig:artefacts}
\end{figure*}

During the training of each segmentation CNN, randomised k-space augmentations affecting image quality are applied on-the-fly. The types of 
augmentation are designed to emulate realistic MRI artefacts and are detailed below. Each augmentation is applied to the k-space by computing the 3D Fast Fourier Transform (FFT) of the input image volume, modifying the k-space, and then computing the inverse 3D FFT, taking the magnitude image scaled between 0 and 1 as input to the network. All augmentations are applied at a rate such that roughly 50\% of images seen by each CNN during training contain artefacts. The order in which k-space augmentations are applied is important to best reflect the MR imaging process, with for instance RF spike $\xrightarrow{}$ noise $\xrightarrow{}$ lowpass filter/k-space sampling. In addition to k-space augmentation, image rotation, scaling and flipping augmentations in all three axes are applied by default to all images, as well as bias field augmentation to account for variation in image intensity across samples.
\\

\noindent \textbf{RF spike artefact} is characterised by dark stripes over the image, as shown in Figure \ref{fig:artefacts} a) 
caused by 
the convolution of spikes in k-space of very high/low intensity 
during the FFT \citep{Zhuo2006}. For 
augmentation we sample uniformly its location in k-space which specifies the angle/frequency of stripes, and its magnitude which defines the intensity of the stripes from between 1 and 10 times the maximum magnitude of the original k-space.
\\

\noindent \textbf{k-Space noise} augmentation involves injecting Gaussian noise into the k-space, as shown in Figure \ref{fig:artefacts} b), to model Rician noise in the image domain. The desired signal-to-noise ratio (SNR) of the image is uniformly sampled between [-10dB, 20dB] and the corresponding amount of complex noise with zero mean and equal variance is added to the k-space.
\\

\noindent \textbf{Blurring artefact} can be observed when acquiring data at lower resolution along an axis prior to resampling.  Low-pass filter applied by truncating the k-space along one or more randomly chosen axes as shown in Figure \ref{fig:artefacts} c) can simulate this effect. 
The width of the filter defines the equivalent downsampling ratio, which is uniformly sampled between 2$\times$ and 12$\times$.
\\

\noindent \textbf{Aliasing/wrap artefact} occurs when the imaging field of view (FOV) is smaller than the anatomy being imaged. This is retrospectively simulated by masking out k-space lines as shown in Figure \ref{fig:artefacts} d). A proportion of k-space lines are either randomly masked uniformly, or at regularly spaced intervals, along a random axis that defines the wraparound direction.
\\

\noindent \textbf{Motion artefact} occurs when the patient moves during the scanning process, and can manifest as blurring, ringing or ghosting effects in the image, depending on the amount and timing of patient movement with regards to the k-space scan trajectory. To simulate motion artefacts, we use the implementation from \cite{shawieee}. The input image volume  is  resampled  according  to  a  randomly  sampled  movement  model,  defined by  a  sequence  of  ‘demeaned’  3D  affine  transformations.  Their respective 3D  Fourier  transforms  are  combined  to  form  a  composite  k-space,  which  is  transformed  back  to the image domain producing the final artefact volume containing motion.
\\

\section{Experimental Setup}
 \begin{figure}[t!]
    \centering
    \includegraphics[width=0.99\textwidth]{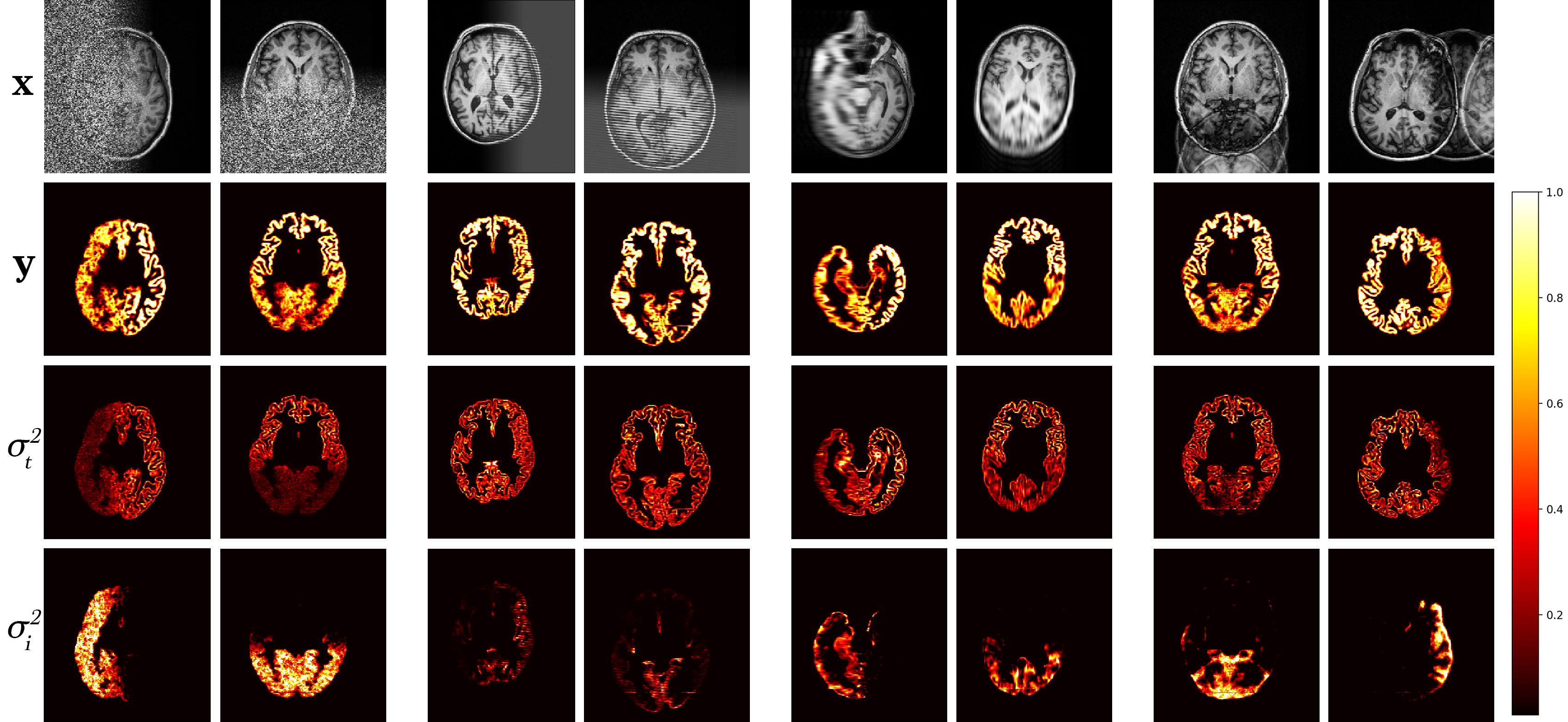}
     \caption{Qualitative results on a hold-out test set of simulated artefacts. First row: artefact-corrupted input image $\mathbf{x}$, second: resulting segmentation $\mathbf{y}$, third: predictive task variance $\sigma^2_t$, and fourth: corresponding augmentation uncertainties $\sigma^2_{i}$ = \{$\sigma^2_{noise}$, $\sigma^2_{rfspike}$, $\sigma^2_{blur}$, $\sigma^2_{wrap}$\}. Best viewed zoomed-in on digital copy.} 
    \label{fig:noise_results}
\end{figure}

\noindent \textbf{Proposed Network Architecture} All CNNs use the updated 3D U-Net architecture from \cite{Isensee2019} as their base architecture implemented in NiftyNet \citep{NiftyNet}. Each network is modified to have two output heads, one for the segmentation prediction $\mathbf{\hat{y}}$ and one for the uncertainty $\sigma^2$, where the uncertainty output from each CNN has a different number of channels -- one for each decoupled uncertainty prediction. We first train the task network to learn a single task uncertainty $\sigma^2_t$. We then train $N$ teacher networks for each augmentation $i$, where each CNN outputs two uncertainties $\sigma^2_t$ and $\sigma^2_i$, corresponding to the task and $i$\textsuperscript{th} augmentation uncertainty respectively. Finally a combined network is trained to learn the task uncertainty and $N$ augmentation uncertainties. Each CNN is trained for 30,000 iterations with a patch size of $96^3$ and batch size of 2 across 4 GPUs with Adam optimiser \citep{Adam} and an initial learning rate of $10^{-4}$. 
\\

\noindent \textbf{Implementation Details} As in \cite{Kendall2017}, for numerical stability,  each CNN is trained to predict log variance $s := \log \sigma^2$ instead of variance $\sigma^2$. In addition, the exponential mapping $\sigma^2 = \exp(s)$ enforces valid positive uncertainty values. We add a small constant $\epsilon$ to the variance to ensure the proper definition of the weighted cross entropy loss,
 $\mathcal{L}_{NN} =
    \mathrm{CE} / (\sigma^2 + \epsilon)
    + \frac{1}{2}
    \log \left( \sigma^2 + \epsilon \right)$.
It's value controls the network's sensitivity to noise and the amount of output uncertainty. For instance, if $\epsilon$ is small, the network is penalised more for making mistakes, outputting higher uncertainty to compensate. If $\epsilon$ is large, the amount the network is penalised is limited by $\mathrm{CE} / \epsilon$ as $\sigma^2 \xrightarrow{} 0$. Furthermore, smaller values of  $\epsilon$ lead to training instability.  Initially, $\epsilon$ is set to $0.05$ and divided by 2 every time the loss plateaus until $\epsilon < 10^{-3}$. In parallel, at each of these steps, the learning rate is also halved.
\newline

\noindent \textbf{Training Data} for this work was obtained from the Alzheimer's Disease Neuroimaging Initiative\footnote[2]{\url{adni.loni.usc.edu}} (ADNI). Launched in 2003, ADNI attempts to assess whether medical imaging and biological markers and clinical assessment can be combined to measure progression of Alzheimer's Disease. For training we use 272 MPRAGE scans that were deemed to be artefact-free, split into 80\% train, 10\% valid and 10\% test. We evaluate our model on simulated and real-world artefacts in the task of grey matter segmentation. \newline

\section{Simulated Data Experiments} A model trained with artefact augmentation was used to perform inference on the hold-out test set. Figure \ref{fig:noise_results} presents a selection of qualitative results. For each image sample, predicted segmentation, task and corresponding augmentation uncertainties are displayed. Areas of high uncertainty are generally in the artefacted regions. This enables us to quickly locate in the volume the image quality issue and judge its effect on the prediction by the level of uncertainty. Note, that in cases of heavy noise, the task uncertainty decreases as the signal is impaired, and therefore the model reverts back to the prior distribution. \newline



\begin{figure*}[!t]
  \centering
  \begin{tabular}{c @{\hspace{0.25em}} c}
    \includegraphics[width=.49\linewidth,trim={0.5cm 0 3.5cm 2cm},clip]{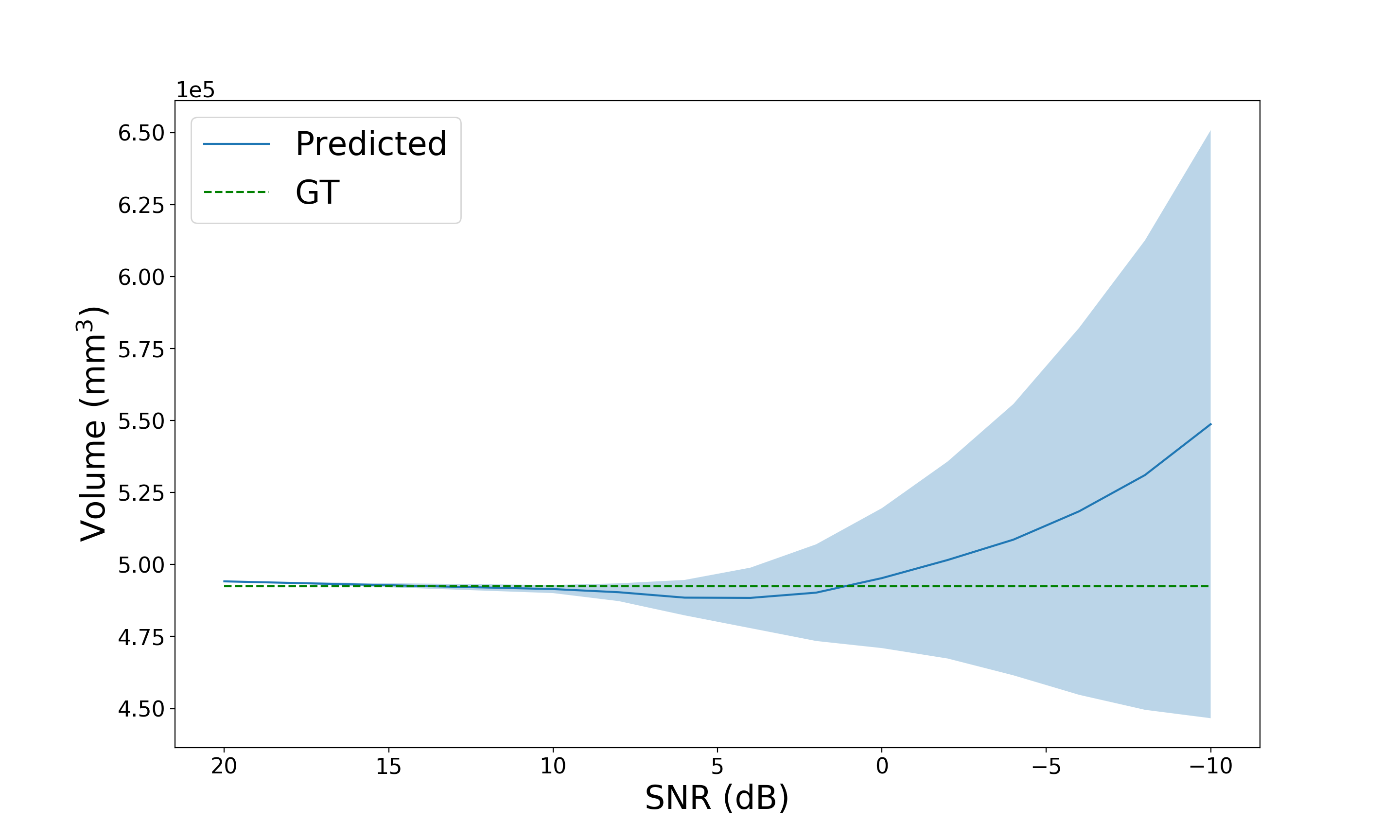} &
    \includegraphics[width=.49\linewidth,trim={0.5cm 0 3.5cm 2cm},clip]{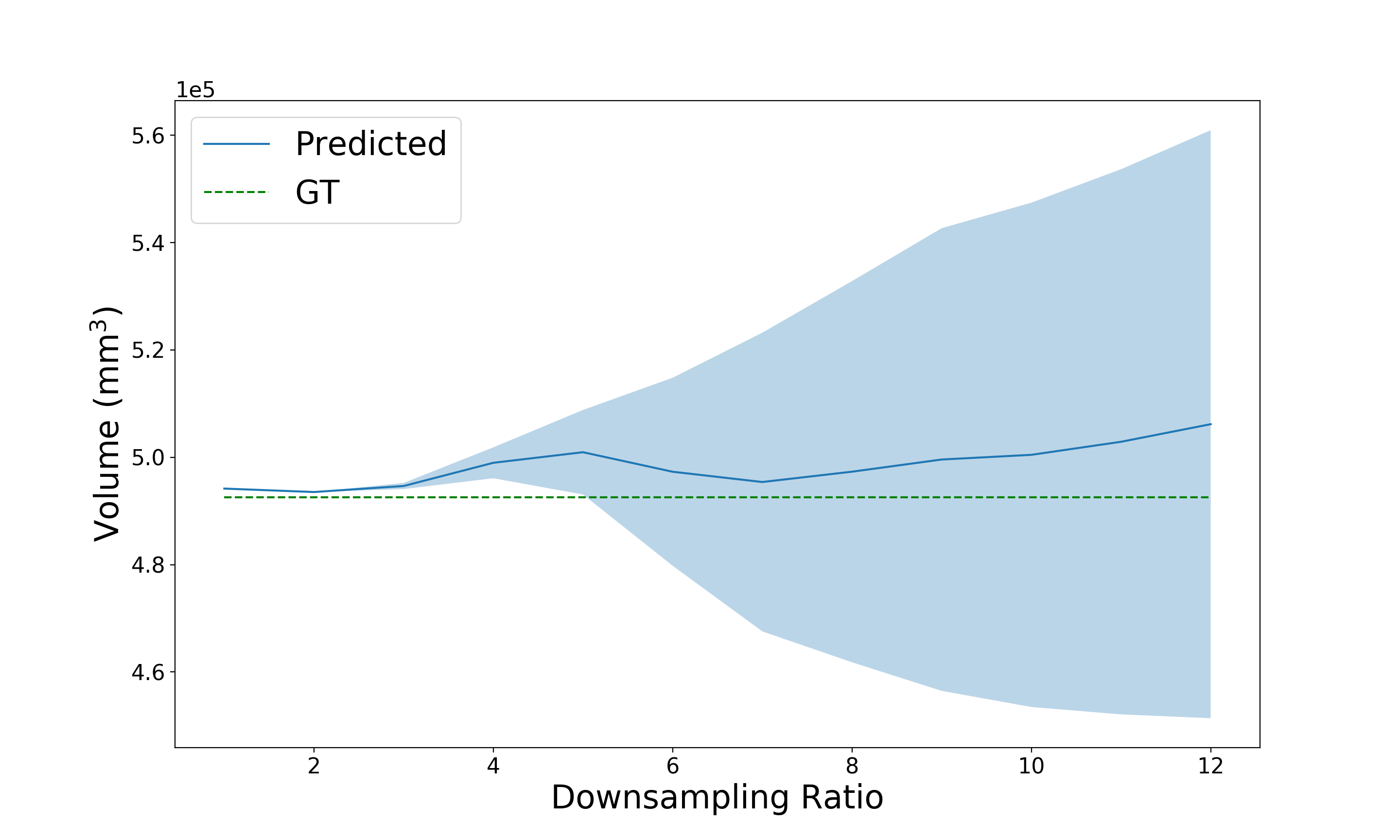} \\
    \small (a) & \small (b)
  \end{tabular}
  \caption{Predicted confidence intervals at $\pm \sigma$ on volume measurements from the grey matter segmentation for images with a) increasing noise and b) increasing blur.}
  \label{fig:errorbars}
\end{figure*}

\noindent \textbf{Entropic Segmentation Uncertainty} The per-voxel variance values predicted by the network pertain to the logit space and are unbounded. While these values directly are useful indicators of uncertainty given the data, they relate to the actual uncertainty of the segmentation prediction through the entropy\footnote[3]{\url{https://github.com/OATML/bdl-benchmarks/tree/alpha/baselines/diabetic_retinopathy_diagnosis}}. Therefore, we can estimate a measure of uncertainty in our probabilities by computing the entropy as $\mathcal{H} = - \sum_{c} p_c \log p_c$,
where $p_c = p(\mathbf{y} = c | \mathbf{f^W}(\mathbf{x}), \sigma)$ are the scaled output logits from the network passed through the Softmax function.
Furthermore, assuming the probabilities are normal distributed (as an approximation), we can use the variance-entropy relation \citep{Jee1986},

\begin{equation}
\mathcal{H} = \frac{1}{2} \ln (2 \pi \mathrm{e} \sigma^2)
\implies
\sigma^2 = \exp (2 \mathcal{H}) / (2 \pi \mathrm{e})
\label{eq:entropyrelation}
\end{equation}

to extract an estimate for the variance $\sigma^2$ of the probability from the entropy. Approximate error bars for the predicted segmentation volume are then obtained by summing the per-voxel variances, resulting in the total variance of the segmentation volume $\pm \sigma$. In Figure \ref{fig:errorbars}, we use our uncertainty predictions to estimate the confidence interval for increasing noise (decreasing SNR) and blurring (increasing downsampling ratio) in the image, relative to clean data. As this is a relative measure of uncertainty, i.e. even noise-free images will have some level of predicted uncertainty, we must transform the uncertainty by some amount to obtain calibrated error bars. In this work we simply compute the difference from the known uncertainty of a noise-free image, but these scaling parameters could be learnt from the data as in \citep{EatonRosen2019}. \newline

\noindent \textbf{Segmentation Quality Estimation}
By posing the issue of MRI quality from the perspective of the downstream task we wish to solve, i.e. a ``task-specific'' notion of quality as opposed to human-perceived ``visual quality,'' we can use the decoupled predictive uncertainties from our model as a measure of segmentation quality. In a real-world data acquisition setting, with the absence of any ground-truth segmentation from a newly scanned patient, clinicians and radiographers would be able to immediately obtain an estimate of the achievable segmentation quality and re-scan the patient if necessary.

 \begin{figure}[tbh!]
    \centering
    \includegraphics[width=0.95\textwidth]{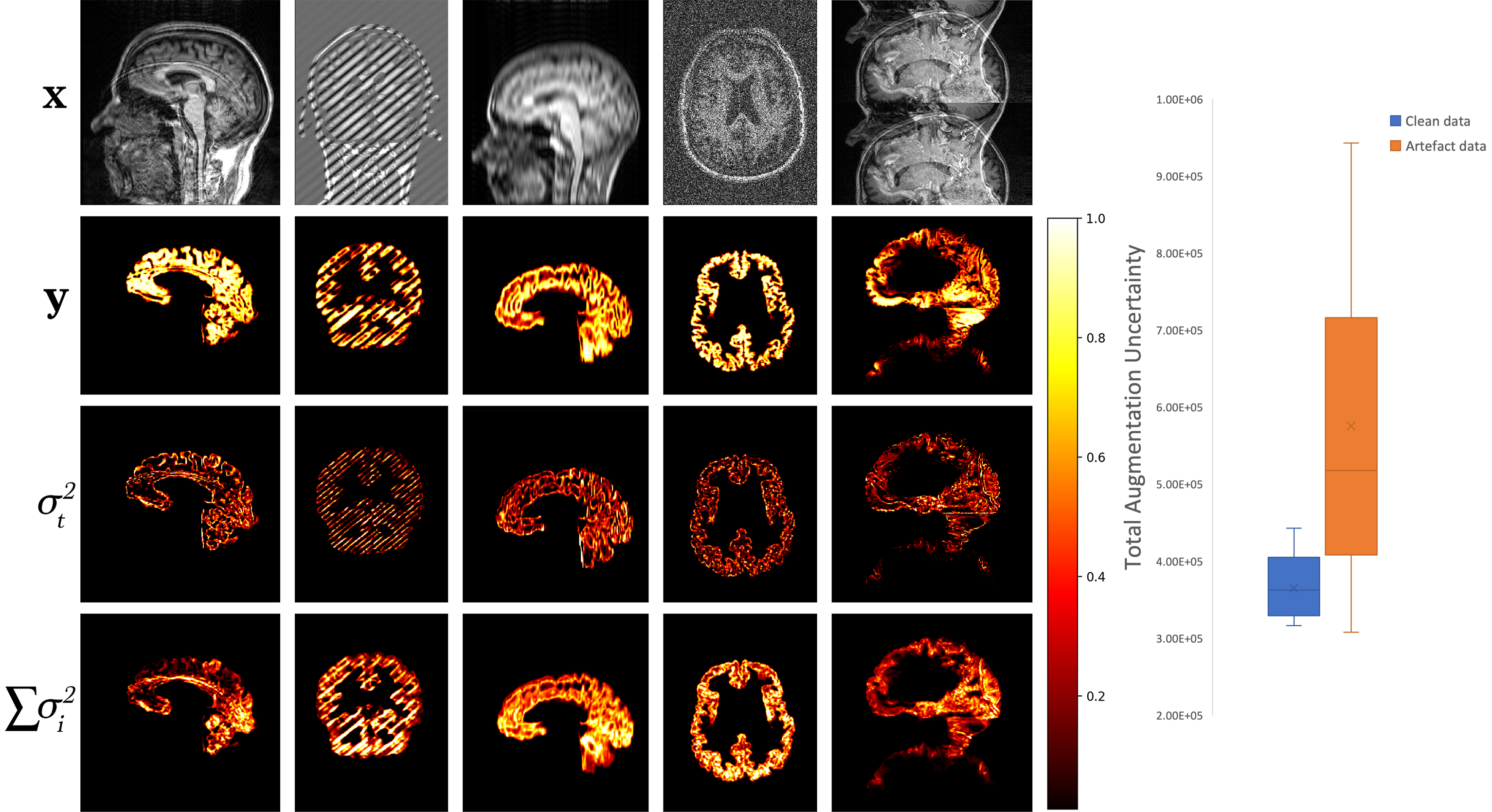}
     \caption{Simulated experiment. Left: samples from the test set. First row: artefact input $\mathbf{x}$, second: predicted segmentation $\mathbf{y}$, third: predictive task variance $\sigma^2_t$, and fourth: total decoupled artefact uncertainty $\sum \sigma^2_i$. Right: total augmentation uncertainty over the datasets.} 
    \label{fig:simulated}
\end{figure}

\begin{figure*}[tbh!]
  \centering
  \begin{tabular}{c @{\hspace{0.25em}} c @{\hspace{0.25em}} c}
    \includegraphics[trim=15 15 10 5,clip,width=.325\linewidth]{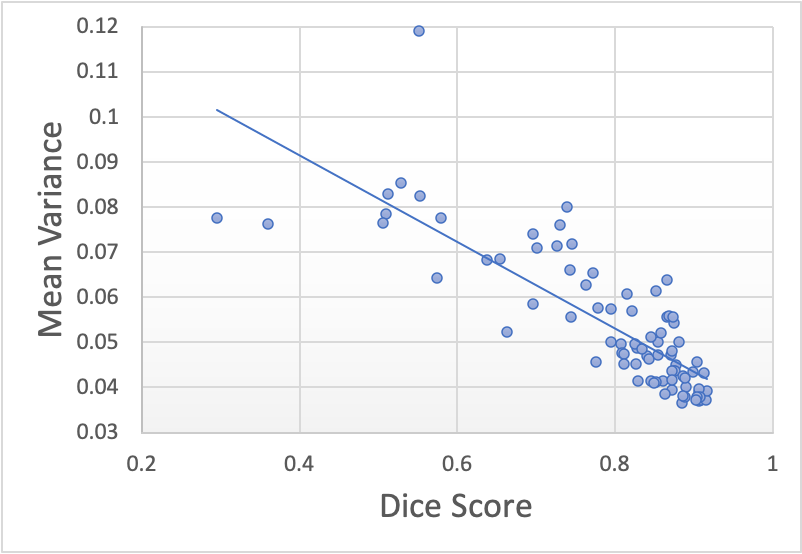} &
    \includegraphics[trim=15 15 10 5,clip,width=.325\linewidth]{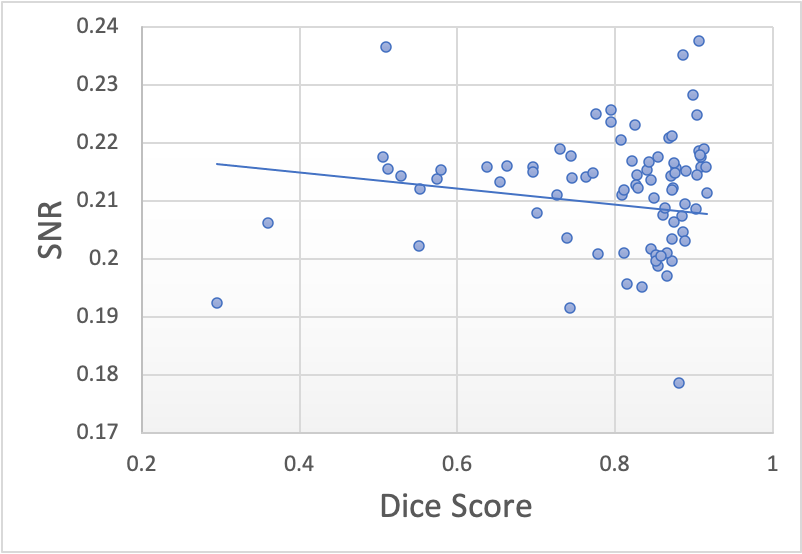} &
    \includegraphics[trim=15 15 10 5,clip,width=.325\linewidth]{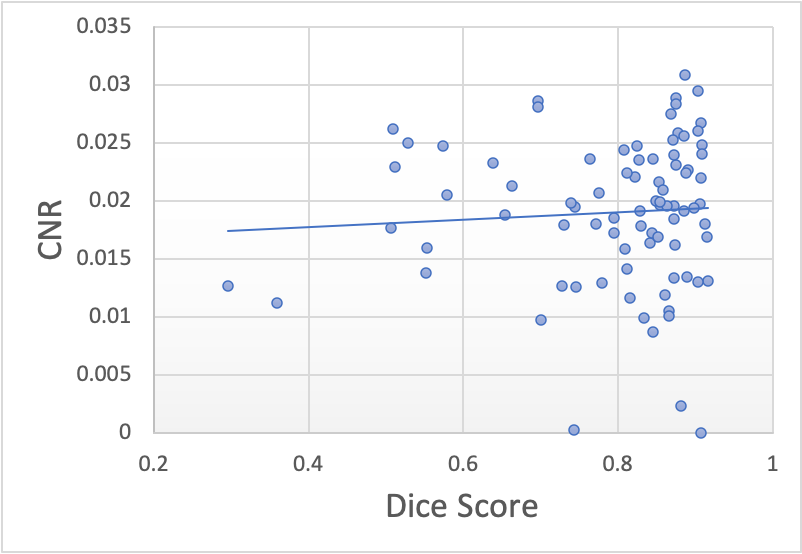}  \\
    \small (a) & \small (b) & \small (c)
  \end{tabular}
  \caption{Simulated experiment: gray matter segmentation Dice scores against different quality metrics, a) uncertainty measured by the mean artefact variance, b) SNR, c) CNR. The decoupled uncertainty best reflects the resulting segmentation quality with a Spearman's rank coefficient of $\rho = -0.850$  and p-value $= 1.395e^{-24}$.}
  \label{fig:dicevar}
\end{figure*}

\begin{table}[tbh!]
\centering
\begin{tabular}{r|c|c|c}
        & $\bar{\sigma}^2_{artefacts}$  & SNR  & CNR  \\ \hline
$\rho$     & -0.850 & 0.326 & 0.200 \\
p-value & $1.395e^{-24}$ & 0.002 & 0.068
\end{tabular}
\caption{Simulated experiment: Spearman's rank correlation between Dice scores and three quality metrics. There is a stronger correlation between the predicted uncertainty from our model to the segmentation Dice scores compared to SNR and CNR.}
\label{table:simulated}
\end{table}

In this experiment, we explore the effectiveness of using the total artefact uncertainty (decoupled from the task uncertainty $\sigma_t^2$) from our model to quantify gray matter segmentation quality when in the presence of image corruption caused by k-space artefacts. We do this by relating the decoupled artefact uncertainty, measured as the mean of the total artefact variance over the image, given by Eq. \ref{eq:meanvar}, to the resulting Dice scores computed between the predicted and ground-truth segmentation maps. Ground-truth segmentation maps are obtained from the clean artefact-free scans using a non-Deep Learning approach \citep{Cardoso2015GeodesicIF}. By simulating k-space artefacts on clean MRI data, we have access to the ground-truth segmentation maps to compute the corresponding Dice scores. We compare the resulting Dice-uncertainty correlation to that from commonly used metrics of MR image quality, including the signal-to-noise ratio (SNR) and contrast-to-noise ratio (CNR). 

\begin{equation}
\bar{\sigma}^2_{artefacts}= \frac{1}{N_{voxels}} \sum_i \sigma^2_i
\label{eq:meanvar}
\end{equation}


The model was trained using the clean ADNI dataset as before, with simulated k-space artefact augmentation applied during training. A test set of artefacts was created using the 10\% hold-out set and simulating k-space artefacts for each unique subject in all 5 artefact sub-type categories described in section 5, resulting in 140 artefact scans. Each artefact scan was generated randomly, thus creating a varied dataset with a range of image degradation severity. 
These artefacts were combined with the clean test images to produce the final test set. 
Inference was performed on the test set, producing 3D segmentation predictions $\mathbf{\hat{y}}$, task uncertainty estimates $\sigma^2_t$, and 5 artefact uncertainty estimates for each of the artefact sub-types $\sigma^2_i$. The decoupled artefact uncertainty maps were summed to produce the total artefact uncertainty decoupled from the task, and the mean variance over the image volume was computed as our metric of segmentation quality. A sample of these results are shown in Figure \ref{fig:simulated}.



For comparison, we compute the SNR and CNR for each scan in the test set. SNR is estimated using Eq. \ref{eq:snr}, where the signal intensity is measured as the mean of the intensity distribution of gray matter, while the noise signal is measured as its standard deviation. 


\begin{equation}
    \mathrm{SNR} = \frac{\mu_{signal}}{\sigma_{noise}} =\frac{\mu_{gray\ matter}}{\sigma_{gray\ matter}}
    \label{eq:snr}
\end{equation}

CNR is computed using Eq. \ref{eq:cnr}, where $\mathrm{C}_{AB}$ is the contrast between tissues $A$ and $B$, estimated by the absolute difference in mean signal intensities $\mu_A$ and  $\mu_B$ from tissues  $A$ and $B$ respectively. For our experiments, contrast is measured between the mean of white and grey matter tissues and their standard deviations provides the noise measurements.

\begin{equation}
    \mathrm{CNR}_{AB} = \frac{\mathrm{C}_{AB}}{\sigma_{noise}}
     = \frac{| \mu_A - \mu_B |}{\sigma_{noise}} = |\mathrm{SNR}_{gray\ matter} - \mathrm{SNR}_{white\ matter}|
    \label{eq:cnr}
\end{equation}

Figure \ref{fig:dicevar} shows the segmentation Dice scores against the three metrics: decoupled artefact uncertainty (mean variance), SNR and CNR. Spearman's rank correlations were computed for each metric and the results are displayed in table \ref{table:simulated}. From the plots we observe that the decoupled artefact uncertainty has the strongest correlation with the resulting Dice scores, with a rank coefficient of $\rho = -0.850$ and a p-value of $1.395e^{-24}$, signifying a very strong correlation. This compares favourably to the SNR metric with $\rho = 0.326$ and p-value of $0.002$ and to CNR with $\rho = 0.200$ and a p-value of $0.068$. Therefore, these results suggest that the mean artefact uncertainty, when decoupled from the task) provided by our decoupled uncertainty model is better correlated with the final segmentation quality than SNR and CNR metrics, and provides a good indicator of segmentation quality.

\newpage
\subsection{Real-world Experiments}

\begin{figure}[tbh!]
    \centering
    \includegraphics[width=0.74\textwidth]{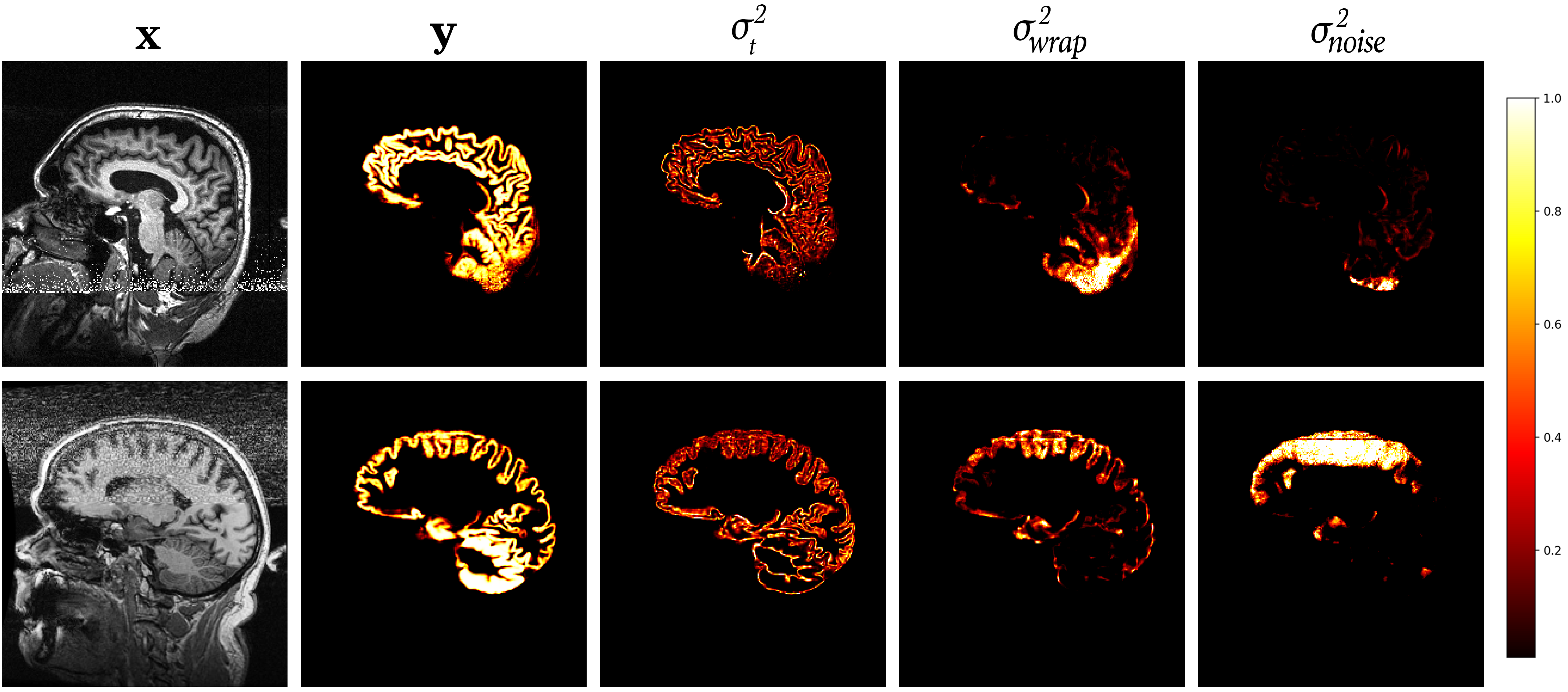}
    \includegraphics[width=0.25\textwidth]{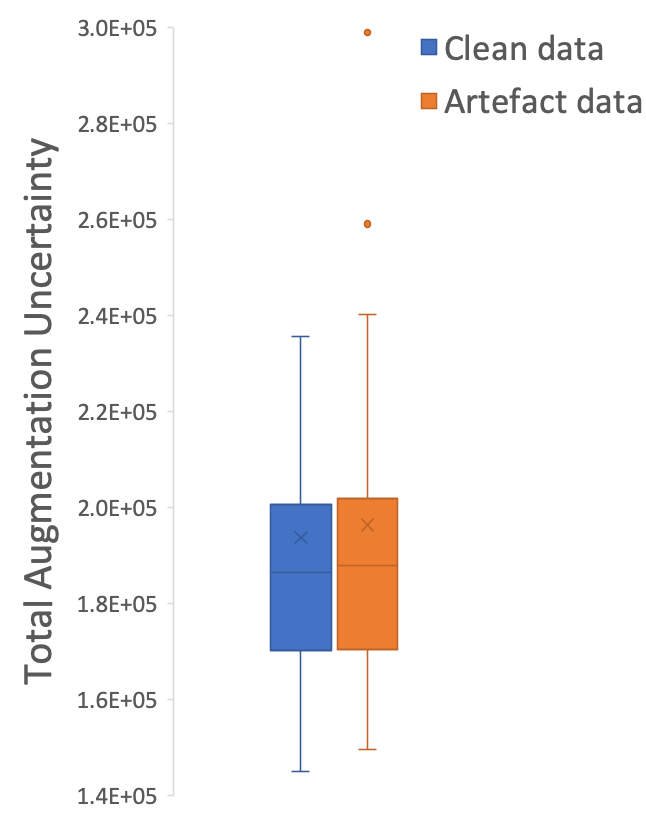}
     \caption{Left: Real-world artefact results. From left to right: artefact input image $\mathbf{x}$, predicted segmentation $\mathbf{y}$, task uncertainty $\sigma^2_t$, and augmentation uncertainties $\sigma^2_{wrap}$ and $\sigma^2_{noise}$.  Right: total augmentation uncertainty over the datasets.} 
    \label{fig:real_results}
\end{figure}

Using the model trained on simulated artefacts, we performed inference on a hold-out dataset of real-world artefact images identified as low quality by expert  raters, predicting segmentation maps $\mathbf{\hat{y}}$, decoupled task uncertainty $\sigma_t^2$ and artefact uncertainties $\sigma_i^2$. A selection of qualitative results are shown in Figure \ref{fig:real_results} (left) for scans identified as containing noise and wrap artefacts. The resulting uncertainty predictions show that the model generalises to real-world artefacts, with generally higher uncertainty in the artefacted regions, immediately highlighting problematic areas in the images. Note, the uncertainty decoupling between artefacts is not perfect and there is some leakage of uncertainty across the outputs.

\begin{figure}[tbh!]
    \centering
    \includegraphics[width=0.7\textwidth]{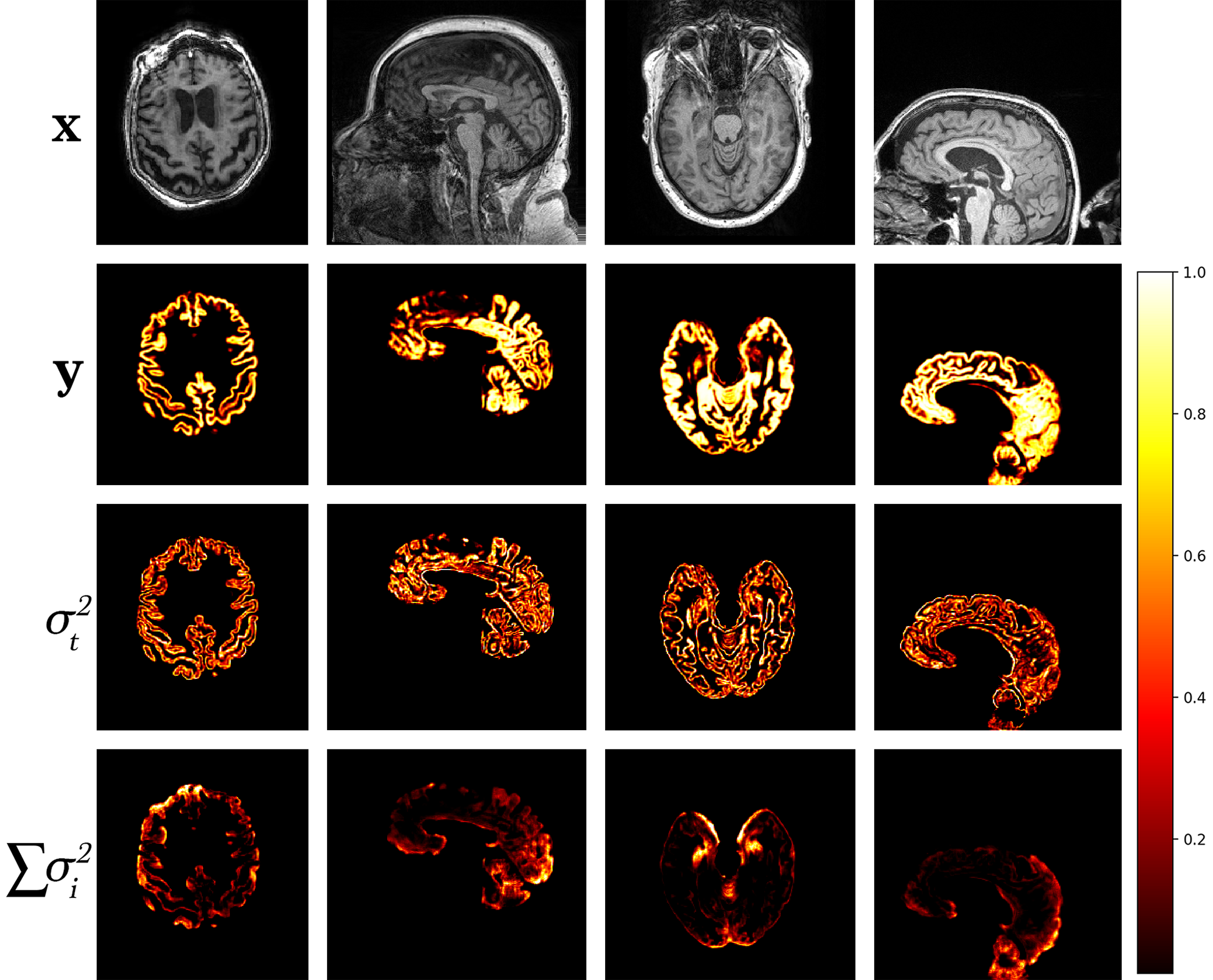}
     \caption{Real-world experiment: hold-out samples of real-world artefacts. First row: artefact input $\mathbf{x}$, second: predicted segmentation $\mathbf{y}$, third: task variance $\sigma^2_t$, and fourth: total decoupled artefact uncertainty $\sum \sigma^2_i$. Best viewed on digital copy.} 
    \label{fig:realworld}
\end{figure}

\begin{figure*}[tbh!]
  \centering
  \begin{tabular}{c @{\hspace{0.25em}} c @{\hspace{0.25em}} c}
    \includegraphics[trim=15 15 10 5,clip,width=.325\linewidth]{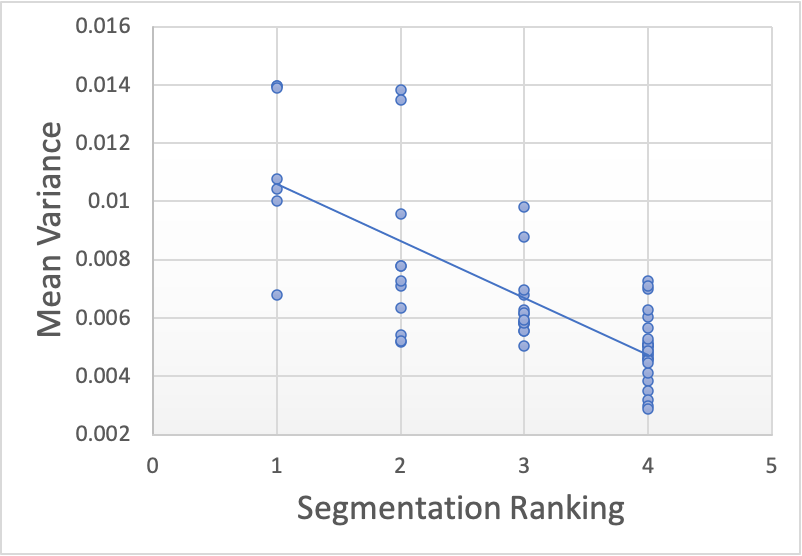} &
    \includegraphics[trim=15 15 10 5,clip,width=.325\linewidth]{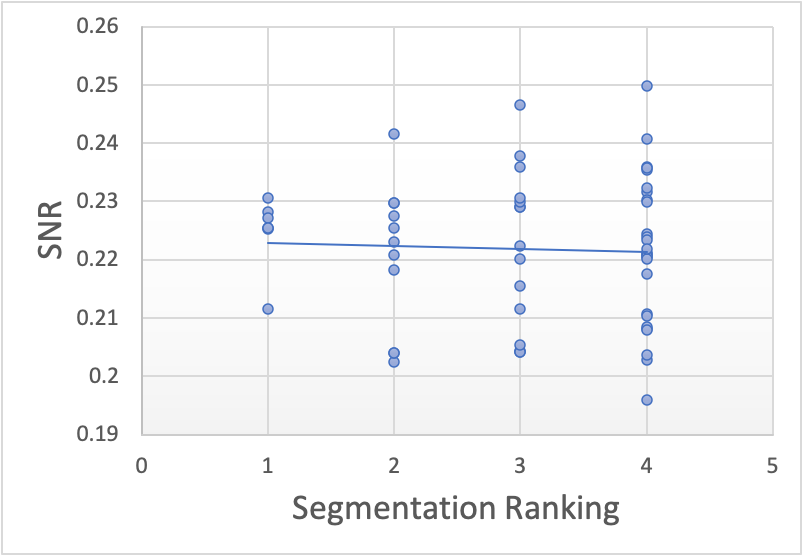} &
    \includegraphics[trim=15 15 10 5,clip,width=.325\linewidth]{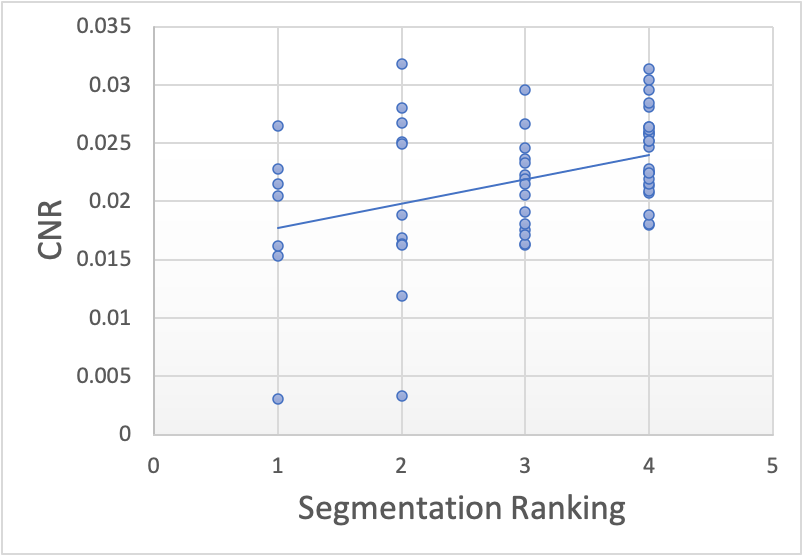}  \\
    \small (a) & \small (b) & \small (c)
  \end{tabular}
  \caption{Real-world experiment: segmentation quality estimated by blindly scoring segmentation maps from 1-4, (1 is worst and 4 is best), vs different quality metrics: a) artefact uncertainty, b) SNR, c) CNR. The uncertainty correlates with segmentation quality with rank coefficient of $\rho = -0.745$ and p-value $= 9.101e^{-12}$.}
  \label{fig:rankvar}
\end{figure*}

Boxplots of total decoupled artefact uncertainty over the clean and artefact datasets are shown in Figure \ref{fig:real_results} (right). Note, the decoupled artefact uncertainty of the artefact set is higher than the clean set, with higher median uncertainty and higher upper and lower quartiles, although the difference is not as significant as one might expect. This is a result of summing many voxels each with small variance, as even when the model is presented with clean images the model trained with log-likelihood loss outputs a small amount of variance per-voxel --- the model cannot output zero variance, so areas of high uncertainty will be found in the tail of the output distribution. This can be partly mitigated by thresholding the variance with a heuristic value, but this would be equivalent to imposing a human-driven quality metric which is inherently subjective. Another reason for the smaller than expected disparity stems from an issue with the labelled data itself. ADNI QC raters label images artefacted if they have an artefact anywhere in the brain, or even outside of the brain, if it is judged to impact downstream brain analysis according to QC protocol\footnote[3]{\url{ https://adni.loni.usc.edu/wp-content/uploads/2010/09/ADNI_MRI_Tech_Proc_Manual.pdf}}, subject to rater interpretation. However, our uncertainty model relates only to the gray matter segmentation region. Thus, if the artefact is not intersecting the gray matter segmentation (or the artefact is in the neck region or outside the skull entirely), the predicted segmentation uncertainty will be low but the image can be still labelled artefacted. This means the segmentation uncertainty can be similar between scans classified as ``clean'' and as ``artefacted.'' Many ADNI scans in the test set deemed unusable have low uncertainty in the brain due to the artefacts being localised outside of the gray matter segmentation, further highlighting the distinction between perceptual quality and algorithmic segmentation quality.
\newline
\newline
\noindent \textbf{Segmentation Quality Estimation}
To further investigate how the model performs on real-world artefacts, we repeated the previous segmentation quality experiment with a separate hold-out test set of 40 scans which had failed manual visual QC inspection by expert raters. A selection of these are shown in Figure \ref{fig:realworld}. The scans failed QC due to a number of reasons including patient motion, image blurring, noise, and wrap artefacts present in the images. Inference was performed on each of the 40 scans using our decoupled segmentation uncertainty model trained with simulated k-space artefacts and the gray matter segmentations and decoupled uncertainty maps were predicted. Since we don't have ground-truth segmentation maps for the real-world artefact data as we had in the simulated experiment, the quality of the predicted segmentation maps were blindly rated (without observing the original image or the network's uncertainty outputs) on scale from 1-4, where 1 is the worst segmentation quality and 4 is the best quality. The resulting segmentation quality scores are plotted against the mean artefact uncertainty, in addition to the estimated SNR and CNR as before. The resulting plots are shown in Figure \ref{fig:rankvar} and the Spearman's rank correlations were computed for each metric and are shown in table \ref{table:real}. Again, we observe that compared to the SNR and CNR metrics, the decoupled artefact variance has the strongest correlation with our estimation of segmentation quality, with a Spearman's rank coefficient of $\rho = -0.745$ and p-value of $p = 9.101e^{-12}$. The uncertainty results show a much stronger correlation to the segmentation quality than SNR or CNR metrics with correlation coefficients of $\rho = -0.045$ and $0.361$ and p-values of $0.730$ and $0.005$ respectively.  Examining the resulting uncertainty maps, we also note that the uncertainty predictions generalise well to out of domain real-world artefacts and that the uncertainty is generally higher in the artefact regions.

\begin{table}[tbh!]
\centering
\begin{tabular}{r|c|c|c}
        & $\bar{\sigma}^2_{artefacts}$  & SNR  & CNR  \\ \hline
$\rho$     & -0.745 & -0.045 & 0.361 \\
p-value & $9.101e^{-12}$ & 0.730 & 0.005
\end{tabular}
\caption{Real-world experiment: Spearman's rank correlation between segmentation quality scores and the mean artefact variance, SNR and CNR. The artefact uncertainty is better correlated to the segmentation quality compared to SNR and CNR.}
\label{table:real}
\end{table}

\section{Discussion}

This work aimed to build a Deep Learning framework capable of identifying and decoupling sources of uncertainty due to MRI artefacts that may affect a given segmentation task. Rather than modelling the perceptual visual quality of MRI scans which is inherently subjective, we have posed the question of scan quality from the perspective of the task. 
We have shown that, using our uncertainty model, it is possible to obtain approximately decoupled uncertainties that, 
reflect the presence (location and severity) of k-space artefacts and that these uncertainties can be used to generate error bars on segmentation measurements. We have also shown that the decoupled artefact uncertainty can be used as a measure of segmentation quality and significantly outperforms the use of SNR and CNR as a metric of segmentation quality, on both simulated and real-world artefact data. \newline

\newpage
\noindent \textbf{Limitations and Future Work}
\vspace{-5pt}

\begin{enumerate}
    \item The model's uncertainty predictions are limited by the fact that we are estimating uncertainty given the data $p(\mathbf{y} | \mathbf{x}, \sigma)$, but we have not modelled the likelihood of the data itself $p(\mathbf{x})$, achievable through methods such as autoencoders. This formulation leads to extrapolation problems on out-of-domain samples. 
    Since the model learns uncertainty from the data in an unsupervised way, we train with the range of artefacts we wish to capture. However, severe image degradation also limits our ability to decouple the task, 
    resulting in an unstable learning process. 
    
    \item For simplicity, sources of uncertainty due to different artefact processes are assumed conditionally independent; however, their interaction is likely to be more complex. 
    Additionally, we assume that the original data is artefact-free, thus artefacts present in the clean data will result in inflated task uncertainty estimates. 
    
    \item 
    Segmentation uncertainty is not a sufficient metric of visual quality. Artefacts outside of the brain are not detected by our model, as the predicted uncertainty relates only to the predicted segmentation. We are exploring the use of Bayesian CNNs for image regression/artefact removal, thus providing a measure of image quality over the image.
    
    \item The ability to decouple uncertainty depends on artefact visual similarity. For instance, blur and motion can be similar, although their appearance in k-space is different -- leveraging k-space information may improve model performance. Also, the more artefact sub-types introduced, the harder it is to decouple due to limited model capacity/increased data complexity. Training also requires a realistic artefact simulator, but 
    real-world data would be preferable. However, gathering paired scans with and without artefacts is a challenge. 
    Attempts to register repeat scans of subjects were made, but 
    imprecise alignment of clean and artefacted scans impacts the uncertainty. 
    
    \item The proposed model is computationally expensive -- a separate 3D CNN is required for each artefact sub-type, with enough features to learn their appearance. 
    With limited GPU resources, patch-based processing can lead to problems with patch border artefacts. Training time is also significant since each CNN must be trained sequentially. 
    
    \item 
    Although developed on a single task, gray matter segmentation, the method has shown transferability from simulated to real data, but not domain transferability. Nonetheless, the model can be extended using Domain Adaptation (adversarial, consistency-based, etc.). The model is also scalable to any segmentation task with arbitrary label complexity as it optimises cross entropy, applying to both binary and multi-class segmentation. Note, assumptions of uncertainty decomposition still hold in this case.
\end{enumerate}

\section{Conclusion}
We have presented a method for estimating  quality-induced task-specific uncertainty using a heteroscedastic noise model. Entirely self-supervised, the proposed model can approximately decouple and localise sources of uncertainty related to different MRI artefacts, thus automatically highlighting problematic areas affecting segmentation predictions. The method is general and may be applied to other automated image analysis processing tasks.


\acks{R. Shaw is funded by an EPSRC CASE studentship. C. H. Sudre was funded by an Alzheimer's Society Fellowship (AS-JF-17-011). M. J. Cardoso was funded by the Wellcome Flagship Programme (WT213038/Z/18/Z) and EPSRC CME (WT203148/Z/16/Z) and the NIHR GSTT BRC. We gratefully acknowledge NVIDIA corporation for the donation of the GPU that was used in the preparation of this work. Data collection and sharing for this project was funded by the Alzheimer's Disease Neuroimaging Initiative (ADNI) (National Institutes of Health Grant U01 AG024904) and DOD ADNI (Department of Defense award number W81XWH-12-2-0012). ADNI is funded by the National Institute on Aging, the National Institute of Biomedical Imaging and Bioengineering, and through generous contributions from the following: AbbVie, Alzheimer's Association; Alzheimer's Drug Discovery Foundation; Araclon Biotech; BioClinica, Inc.; Biogen; Bristol-Myers Squibb Company; CereSpir, Inc.; Cogstate; Eisai Inc.; Elan Pharmaceuticals, Inc.; Eli Lilly and Company; EuroImmun; F. Hoffmann-La Roche Ltd and its affiliated company Genentech, Inc.; Fujirebio; GE Healthcare; IXICO Ltd.; Janssen Alzheimer Immunotherapy Research \& Development, LLC.; Johnson \& Johnson Pharmaceutical Research \& Development LLC.; Lumosity; Lundbeck; Merck \& Co., Inc.; Meso Scale Diagnostics, LLC.; NeuroRx Research; Neurotrack Technologies; Novartis Pharmaceuticals Corporation; Pfizer Inc.; Piramal Imaging; Servier; Takeda Pharmaceutical Company; and Transition Therapeutics. The Canadian Institutes of Health Research is providing funds to support ADNI clinical sites in Canada. Private sector contributions are facilitated by the Foundation for the National Institutes of Health (www.fnih.org). The grantee organization is the Northern California Institute for Research and Education, and the study is coordinated by the Alzheimer's Therapeutic Research Institute at the University of Southern California. ADNI data are disseminated by the Laboratory for Neuro Imaging at the University of Southern California.}

%
\ethics{The work follows appropriate ethical standards in conducting research and writing the manuscript, following all applicable laws and regulations regarding treatment of animals or human subjects.}

\coi{We declare we don't have conflicts of interest.}

\bibliography{melba-sample}

\begin{thebibliography}{41}
\providecommand{\natexlab}[1]{#1}
\providecommand{\url}[1]{\texttt{#1}}
\expandafter\ifx\csname urlstyle\endcsname\relax
  \providecommand{\doi}[1]{doi: #1}\else
  \providecommand{\doi}{doi: \begingroup \urlstyle{rm}\Url}\fi

\bibitem[Bragman et~al.(2018)Bragman, Tanno, Eaton-Rosen, Li, Hawkes,
  et~al.]{Bragman2018}
Felix J.~S. Bragman, Ryutaro Tanno, Zach Eaton-Rosen, Wenqi Li, David~J.
  Hawkes, et~al.
\newblock Uncertainty in multitask learning: joint representations for
  probabilistic mr-only radiotherapy planning.
\newblock In \emph{MICCAI}, 2018.

\bibitem[Brusini et~al.(2020)Brusini, Padilla, Barroso, Skoog, Smedby, Westman,
  and Wang]{Brusini2020ADL}
Irene Brusini, D.~Padilla, J.~Barroso, I.~Skoog, Orjan Smedby, E.~Westman, and
  C.~Wang.
\newblock A deep learning-based pipeline for error detection and quality
  control of brain mri segmentation results.
\newblock \emph{arXiv: Image and Video Processing}, 2020.

\bibitem[Cardoso et~al.(2015)Cardoso, Modat, Wolz, Melbourne, Cash, Rueckert,
  and Ourselin]{Cardoso2015GeodesicIF}
M.~J. Cardoso, M.~Modat, R.~Wolz, A.~Melbourne, D.~Cash, D.~Rueckert, and
  S.~Ourselin.
\newblock Geodesic information flows: Spatially-variant graphs and their
  application to segmentation and fusion.
\newblock \emph{IEEE Transactions on Medical Imaging}, 34:\penalty0 1976--1988,
  2015.

\bibitem[Combalia et~al.(2020)Combalia, Hueto, Puig, Malvehy, and
  Vilaplana]{Combalia2020Uncertainty}
M.~Combalia, Ferran Hueto, S.~Puig, J.~Malvehy, and Ver{\'o}nica Vilaplana.
\newblock Uncertainty estimation in deep neural networks for dermoscopic image
  classification.
\newblock \emph{2020 IEEE/CVF Conference on Computer Vision and Pattern
  Recognition Workshops (CVPRW)}, pages 3211--3220, 2020.

\bibitem[Cubuk et~al.(2019)Cubuk, Zoph, Shlens, and Le]{RandAugment}
E.~Cubuk, Barret Zoph, Jonathon Shlens, and Quoc~V. Le.
\newblock Randaugment: Practical data augmentation with no separate search.
\newblock \emph{ArXiv}, abs/1909.13719, 2019.

\bibitem[Devries and Taylor(2018)]{Devries2018LeveragingUE}
Terrance Devries and Graham~W. Taylor.
\newblock Leveraging uncertainty estimates for predicting segmentation quality.
\newblock \emph{ArXiv}, abs/1807.00502, 2018.

\bibitem[Eaton-Rosen et~al.(2019)Eaton-Rosen, Varsavsky, Ourselin, and
  Cardoso]{EatonRosen2019}
Zach Eaton-Rosen, Thomas Varsavsky, S{\'e}bastien Ourselin, and M.~Jorge
  Cardoso.
\newblock As easy as 1, 2... 4? uncertainty in counting tasks for medical
  imaging.
\newblock In \emph{MICCAI}, 2019.

\bibitem[Esteban et~al.(2017)Esteban, Birman, Schaer, Koyejo, Poldrack, and
  Gorgolewski]{Esteban2017MRIQCAT}
Oscar Esteban, Daniel Birman, M.~Schaer, Oluwasanmi~O. Koyejo, R.~Poldrack, and
  Krzysztof~J. Gorgolewski.
\newblock Mriqc: Advancing the automatic prediction of image quality in mri
  from unseen sites.
\newblock \emph{PLoS ONE}, 12, 2017.

\bibitem[Gal and Ghahramani(2016)]{Gal2016Dropout}
Yarin Gal and Zoubin Ghahramani.
\newblock Dropout as a bayesian approximation: Representing model uncertainty
  in deep learning.
\newblock \emph{ArXiv}, abs/1506.02142, 2016.

\bibitem[Gedamu et~al.(2008)Gedamu, Collins, and Arnold]{Elias}
Elias Gedamu, D.~Louis Collins, and Douglas~L. Arnold.
\newblock Automated quality control of brain mr images.
\newblock \emph{Journal of Magnetic Resonance Imaging}, 28\penalty0
  (2):\penalty0 308--319, 2008.

\bibitem[Gibson et~al.(2017)]{NiftyNet}
Eli Gibson et~al.
\newblock Niftynet: a deep-learning platform for medical imaging.
\newblock \emph{CoRR}, 2017.
\newblock URL \url{http://arxiv.org/abs/1709.03485}.

\bibitem[Graham et~al.(2018)Graham, Drobnjak, and Zhang]{Graham2018}
Mark~S. Graham, Ivana Drobnjak, and Hui Zhang.
\newblock A supervised learning approach for diffusion mri quality control with
  minimal training data.
\newblock \emph{NeuroImage}, 178:\penalty0 668--676, 2018.

\bibitem[Hendrycks et~al.(2020)Hendrycks, Mu, Cubuk, Zoph, Gilmer, and
  Lakshminarayanan]{Hendrycks2020AugMixAS}
Dan Hendrycks, Norman Mu, E.~D. Cubuk, Barret Zoph, J.~Gilmer, and Balaji
  Lakshminarayanan.
\newblock Augmix: A simple data processing method to improve robustness and
  uncertainty.
\newblock \emph{ArXiv}, abs/1912.02781, 2020.

\bibitem[Hinton et~al.(2015)Hinton, Vinyals, and Dean]{Hinton2015}
Geoffrey~E. Hinton, Oriol Vinyals, and Jeffrey Dean.
\newblock Distilling the knowledge in a neural network.
\newblock \emph{ArXiv}, abs/1503.02531, 2015.

\bibitem[Hu et~al.(2019)Hu, Worrall, Knegt, Veeling, Huisman, and
  Welling]{Hu2019Supervised}
Shi Hu, Daniel Worrall, Stefan J.~L. Knegt, Bas Veeling, H.~Huisman, and
  M.~Welling.
\newblock Supervised uncertainty quantification for segmentation with multiple
  annotations.
\newblock In \emph{MICCAI}, 2019.

\bibitem[Isensee et~al.(2019)Isensee, Petersen, Klein, Zimmerer, Jaeger,
  et~al.]{Isensee2019}
Fabian Isensee, Jens Petersen, Andr{\'e} Klein, David Zimmerer, Paul~F. Jaeger,
  et~al.
\newblock Abstract: nnu-net: Self-adapting framework for u-net-based medical
  image segmentation.
\newblock In \emph{Bildverarbeitung f{\"u}r die Medizin}, 2019.

\bibitem[Isola et~al.(2017)Isola, Zhu, Zhou, and
  Efros]{Isola2017ImagetoImageTW}
Phillip Isola, Jun-Yan Zhu, Tinghui Zhou, and Alexei~A. Efros.
\newblock Image-to-image translation with conditional adversarial networks.
\newblock \emph{2017 IEEE Conference on Computer Vision and Pattern Recognition
  (CVPR)}, pages 5967--5976, 2017.

\bibitem[Jee and Ratnaparkhi(1986)]{Jee1986}
Debabrata~Mukher Jee and Makarand~V Ratnaparkhi.
\newblock On the functional relationship between entropy and variance with
  related applications.
\newblock \emph{Communications in Statistics - Theory and Methods}, 15\penalty0
  (1):\penalty0 291--311, 1986.

\bibitem[Kendall and Gal(2017)]{Kendall2}
Alex Kendall and Yarin Gal.
\newblock What uncertainties do we need in bayesian deep learning for computer
  vision?
\newblock In \emph{NIPS}, 2017.

\bibitem[Kendall et~al.(2017)Kendall, Gal, and Cipolla]{Kendall2017}
Alex Kendall, Yarin Gal, and Roberto Cipolla.
\newblock Multi-task learning using uncertainty to weigh losses for scene
  geometry and semantics.
\newblock \emph{CVPR}, pages 7482--7491, 2017.

\bibitem[Kingma and Ba(2015)]{Adam}
Diederik~P. Kingma and Jimmy Ba.
\newblock Adam: {A} method for stochastic optimization.
\newblock In \emph{ICLR}, 2015.
\newblock URL \url{http://arxiv.org/abs/1412.6980}.

\bibitem[Kohl et~al.(2018)Kohl, Romera-Paredes, Meyer, Fauw, Ledsam,
  Maier-Hein, Eslami, Rezende, and Ronneberger]{Kohl2018APU}
Simon A.~A. Kohl, B.~Romera-Paredes, Clemens Meyer, J.~Fauw, Joseph~R. Ledsam,
  Klaus Maier-Hein, S.~Eslami, Danilo~Jimenez Rezende, and O.~Ronneberger.
\newblock A probabilistic u-net for segmentation of ambiguous images.
\newblock In \emph{NeurIPS}, 2018.

\bibitem[Lakshminarayanan et~al.(2017)Lakshminarayanan, Pritzel, and
  Blundell]{Lak2017}
Balaji Lakshminarayanan, A.~Pritzel, and Charles Blundell.
\newblock Simple and scalable predictive uncertainty estimation using deep
  ensembles.
\newblock \emph{ArXiv}, abs/1612.01474, 2017.

\bibitem[Muelly et~al.(2017)Muelly, B.~Stoddard, and S.~Vasanwala]{Muelly}
Michael Muelly, Paul B.~Stoddard, and Shreyas S.~Vasanwala.
\newblock Automated quality control of mr images using deep convolutional
  neural networks.
\newblock 2017.

\bibitem[Prado et~al.(2019)Prado, Kausik, and Venkataramanan]{Prado2019DualNN}
Augustin Prado, Ravinath Kausik, and Lalitha Venkataramanan.
\newblock Dual neural network architecture for determining epistemic and
  aleatoric uncertainties.
\newblock \emph{ArXiv}, abs/1910.06153, 2019.

\bibitem[Roy et~al.(2019)Roy, Conjeti, Navab, and Wachinger]{Roy2019BayesianQM}
Abhijit~Guha Roy, Sailesh Conjeti, Nassir Navab, and C.~Wachinger.
\newblock Bayesian quicknat: Model uncertainty in deep whole-brain segmentation
  for structure-wise quality control.
\newblock \emph{NeuroImage}, 195:\penalty0 11--22, 2019.

\bibitem[Sadri et~al.(2020)Sadri, Janowczyk, Zou, Verma, Antunes, Madabhushi,
  Tiwari, and Viswanath]{Sadri2020MRQyAO}
A.~Sadri, Andrew Janowczyk, R.~Zou, R.~Verma, Jacob Antunes, A.~Madabhushi,
  P.~Tiwari, and S.~Viswanath.
\newblock Mrqy: An open-source tool for quality control of mr imaging data.
\newblock \emph{ArXiv}, abs/2004.04871, 2020.

\bibitem[Shaw et~al.(2020)Shaw, Sudre, Ourselin, and Cardoso]{Shaw2020AHU}
R.~Shaw, C.~Sudre, S{\'e}bastien Ourselin, and M.~Cardoso.
\newblock A heteroscedastic uncertainty model for decoupling sources of mri
  image quality.
\newblock In \emph{MIDL}, 2020.

\bibitem[{Shaw} et~al.(2020){Shaw}, {Sudre}, {Varsavsky}, {Ourselin}, and
  {Cardoso}]{shawieee}
R.~{Shaw}, C.~H. {Sudre}, T.~{Varsavsky}, S.~{Ourselin}, and M.~J. {Cardoso}.
\newblock A k-space model of movement artefacts: Application to segmentation
  augmentation and artefact removal.
\newblock \emph{IEEE Transactions on Medical Imaging}, pages 1--1, 2020.

\bibitem[Sudre et~al.(2019)Sudre, Anson, Ingala, Lane, Jimenez,
  et~al.]{Sudre19}
Carole~H. Sudre, Beatriz~Gomez Anson, Silvia Ingala, Chris~D. Lane, Daniel
  Jimenez, et~al.
\newblock Let's agree to disagree: Learning highly debatable multirater
  labelling.
\newblock In \emph{MICCAI}, pages 665--673, 2019.

\bibitem[Sun et~al.(2014)Sun, Barnes, Dowling, Menk, Stanwell, and Greer]{Sun}
Jidi Sun, Michael Barnes, Jason Dowling, Frederick Menk, Peter Stanwell, and
  Peter Greer.
\newblock An open source automatic quality assurance (osaqa) tool for the acr
  mri phantom.
\newblock \emph{Australasian physical \& engineering sciences in medicine /
  supported by the Australasian College of Physical Scientists in Medicine and
  the Australasian Association of Physical Sciences in Medicine}, 11 2014.
\newblock \doi{10.1007/s13246-014-0311-8}.

\bibitem[Tanno et~al.(2019)Tanno, Worrall, Kaden, Ghosh, Grussu,
  et~al.]{Tanno2019UncertaintyQI}
Ryutaro Tanno, Daniel Worrall, Enrico Kaden, Aurobrata Ghosh, Francesco Grussu,
  et~al.
\newblock Uncertainty quantification in deep learning for safer neuroimage
  enhancement.
\newblock \emph{ArXiv}, 2019.
\newblock URL \url{http://arxiv.org/abs/1907.13418}.

\bibitem[Tarvainen and Valpola(2017)]{Tarvainen2017}
Antti Tarvainen and Harri Valpola.
\newblock Mean teachers are better role models: Weight-averaged consistency
  targets improve semi-supervised deep learning results.
\newblock In \emph{NIPS}, 2017.

\bibitem[Wang et~al.(2018)Wang, Li, Aertsen, Deprest, Ourselin,
  et~al.]{Wang2018}
Guotai Wang, Wenqi Li, Michael Aertsen, Jan Deprest, S{\'e}bastien Ourselin,
  et~al.
\newblock Aleatoric uncertainty estimation with test-time augmentation for
  medical image segmentation with convolutional neural networks.
\newblock In \emph{Neurocomputing}, 2018.

\bibitem[Wang et~al.(2020)Wang, Tarroni, Qin, Mo, Dai, Chen, Glocker, Guo,
  Rueckert, and Bai]{Wang2020DeepGM}
S.~Wang, G.~Tarroni, Chen Qin, Y.~Mo, Chengliang Dai, Chen Chen, B.~Glocker,
  Y.~Guo, D.~Rueckert, and Wenjia Bai.
\newblock Deep generative model-based quality control for cardiac mri
  segmentation.
\newblock \emph{ArXiv}, abs/2006.13379, 2020.

\bibitem[Xie et~al.(2019)Xie, Hovy, Luong, and Le]{NoisyStudent}
Qizhe Xie, Eduard Hovy, Minh-Thang Luong, and Quoc~V. Le.
\newblock Self-training with noisy student improves imagenet classification.
\newblock \emph{arXiv preprint arXiv:1911.04252}, 2019.

\bibitem[Xie et~al.(2020)Xie, Hovy, Luong, and Le]{Xie2020SelfTrainingWN}
Qizhe Xie, E.~Hovy, Minh-Thang Luong, and Quoc~V. Le.
\newblock Self-training with noisy student improves imagenet classification.
\newblock \emph{2020 IEEE/CVF Conference on Computer Vision and Pattern
  Recognition (CVPR)}, pages 10684--10695, 2020.

\bibitem[Yun et~al.(2019)Yun, Han, Oh, Chun, Choe, and Yoo]{Yun2019CutMixRS}
Sangdoo Yun, Dongyoon Han, Seong~Joon Oh, Sanghyuk Chun, Junsuk Choe, and
  Youngjoon Yoo.
\newblock Cutmix: Regularization strategy to train strong classifiers with
  localizable features.
\newblock \emph{2019 IEEE/CVF International Conference on Computer Vision
  (ICCV)}, pages 6022--6031, 2019.

\bibitem[Zhang et~al.(2018)Zhang, Ciss{\'e}, Dauphin, and
  Lopez-Paz]{Zhang2018mixupBE}
Hongyi Zhang, M.~Ciss{\'e}, Yann Dauphin, and David Lopez-Paz.
\newblock mixup: Beyond empirical risk minimization.
\newblock \emph{ArXiv}, abs/1710.09412, 2018.

\bibitem[Zhao et~al.(2017)Zhao, Gallo, Frosio, and Kautz]{Zhao2017LossFF}
Hang Zhao, Orazio Gallo, Iuri Frosio, and Jan Kautz.
\newblock Loss functions for image restoration with neural networks.
\newblock \emph{IEEE Transactions on Computational Imaging}, 3:\penalty0
  47--57, 2017.

\bibitem[Zhuo and Gullapalli(2006)]{Zhuo2006}
Jiachen Zhuo and Rao~P. Gullapalli.
\newblock Aapm/rsna physics tutorial for residents: Mr artifacts, safety, and
  quality control.
\newblock \emph{Radiographics}, 26 1:\penalty0 275--97, 2006.

\end{thebibliography}


\end{document}